\documentclass{vldb}
\usepackage{graphicx}
\usepackage{balance}  % for  \balance command ON LAST PAGE  (only there!)
\usepackage{enumitem}
\usepackage{fancyvrb}
\usepackage{tabularx}
\usepackage{multirow}
\usepackage[font=normalsize,labelfont=bf]{caption}

\newcommand{\refsec}[1]{Section~\ref{#1}}

\newcommand{\reffig}[1]{Figure~\ref{#1}}

\newcommand{\apply}{\mathcal{A}}
\newcommand{\mysubsection}[1]{\vspace{3mm} \par\noindent{\bf #1}:}

\newcommand{\onecolfigure}[3]{
	\begin{figure}[tbp]
		\centering
		\centerline{\epsfxsize=\columnwidth \epsffile{#1}}
		\vspace*{0.2em}
		\centerline{\parbox{\columnwidth}{\caption{#2} \label{#3}}}
		\vspace*{-1.7em}
	\end{figure}}
	
\newcommand{\onecolfigurenogap}[3]{
	\begin{figure}[tbp]
		\centering
		\vspace*{-0.7em}
		\centerline{\epsfxsize=\columnwidth \epsffile{#1}}
		\vspace*{-1em}
		\centerline{\parbox{\columnwidth}{\caption{#2} \label{#3}}}
		\vspace*{-2.5em}
	\end{figure}}
		
\newcommand{\twocolfigure}[3]{
	\begin{figure*}[tb]
		\centering
		\centerline{\epsfxsize=2.1\columnwidth \epsffile{#1}}
		\vspace*{.3em}
		\centerline{\parbox{2.1\columnwidth}{\caption{#2} \label{#3}}}
		\vspace*{-2em}
	\end{figure*}}

\newcommand{\divider}[0]{
	\vspace{-3mm}
	\noindent\rule{80mm}{.1pt}
	\vspace{-2mm}}

\begin{document}

% ****************** TITLE ****************************************

%\title{A Sample {\ttlit Proceedings of the VLDB Endowment} Paper in LaTeX
%Format\titlenote{for use with vldb.cls}}
\title{Optimization of Imperative Programs in a Relational Database\titlenote{Extended version of the paper titled ``Froid: Optimization of Imperative Programs in a Relational Database'' in PVLDB 11(4), 2017. DOI: 10.1145/3164135.3164140.~\cite{FroidVldb}}}
	
\subtitle{Technical Report}
% possible, but not really needed or used for PVLDB:
%\subtitle{[Extended Abstract]
%\titlenote{A full version of this paper is available as\textit{Author's Guide to Preparing ACM SIG Proceedings Using \LaTeX$2_\epsilon$\ and BibTeX} at \texttt{www.acm.org/eaddress.htm}}}

% ****************** AUTHORS **************************************


\numberofauthors{3} %  in this sample file, there are a *total*

\author{
% 1st. author
	\alignauthor
	Karthik Ramachandra\\
	\affaddr{Microsoft Gray Systems Lab}\\
	%\affaddr{1932 Wallamaloo Lane}\\
	%\affaddr{Wallamaloo, New Zealand}\\
	\email{\small{karam@microsoft.com}}	
	% 2nd. author
	\alignauthor
	Kwanghyun Park\\
	\affaddr{Microsoft Gray Systems Lab}\\
	%\affaddr{P.O. Box 1212}\\
	%\affaddr{Dublin, Ohio 43017-6221}\\
	\email{\small{kwpark@microsoft.com}}	
	% 3rd. author
	\alignauthor K. Venkatesh Emani\titlenote{Work done as an intern at Microsoft Gray Systems Lab}\\
	\affaddr{IIT Bombay}\\
	%\affaddr{1 Th{\o}rv{\"a}ld Circle}\\
	%\affaddr{Hekla, Iceland}\\
	\email{\small{venkateshek@cse.iitb.ac.in}}	
	\and  % use '\and' if you need 'another row' of author names	
	% 4th. author
	\alignauthor Alan Halverson\\
	\affaddr{Microsoft Gray Systems Lab}\\
	%\affaddr{Brookhaven National Lab}\\
	%\affaddr{P.O. Box 5000}\\
	\email{\small{alanhal@microsoft.com}}	
	% 5th. author
	\alignauthor C\'esar Galindo-Legaria\\
	\affaddr{Microsoft}\\
	%\affaddr{Moffett Field}\\
	%\affaddr{California 94035}\\
	\email{\small{cesarg@microsoft.com}}	
	% 6th. author
	\alignauthor Conor Cunningham\\
	\affaddr{Microsoft}\\
	%\affaddr{8600 Datapoint Drive}\\
	%\affaddr{San Antonio, Texas 78229}\\
	\email{\small{conorc@microsoft.com}}
}
% There's nothing stopping you putting the seventh, eighth, etc.
% author on the opening page (as the 'third row') but we ask,
% for aesthetic reasons that you place these 'additional authors'
% in the \additional authors block, viz.
%\additionalauthors{Additional authors: John Smith (The Th{\o}rv\"{a}ld Group, {\texttt{jsmith@affiliation.org}}), Julius P.~Kumquat
%(The \raggedright{Kumquat} Consortium, {\small \texttt{jpkumquat@consortium.net}}), and Ahmet Sacan (Drexel University, {\small \texttt{ahmetdevel@gmail.com}})}
% Just remember to make sure that the TOTAL number of authors
% is the number that will appear on the first page PLUS the
% number that will appear in the \additionalauthors section.

\maketitle
%\graphicspath{{../}}
\begin{abstract}
%Relational databases have focused primarily on efficient evaluation of 
%declarative SQL. Although imperative functions and procedures are supported, 
%they are evaluated in a na\"{\i}ve and highly inefficient manner. Functions 
%provide a powerful abstraction to achieve modularity and code reuse, and are 
%often a preferred way to express intent in many applications. Unfortunately, 
%their abysmal performance discourages, and often prohibits their use. In this 
%work, we address this important problem that has hitherto received little 
%attention.

For decades, RDBMSs have supported declarative SQL as well as imperative 
functions and procedures as ways for users to express data processing tasks. 
While the evaluation of declarative SQL has received a lot of attention 
resulting in highly sophisticated techniques, the evaluation of 
imperative programs has remained na\"{\i}ve and highly inefficient. Imperative programs 
offer several benefits over SQL and hence are often preferred and widely used. But unfortunately, their abysmal performance 
discourages, and even prohibits their use in many situations. 
We address this important problem that has hitherto received little 
attention.

We present Froid, an extensible framework for optimizing imperative programs in 
relational databases. Froid's novel approach automatically transforms entire User Defined 
Functions (UDFs) into relational algebraic expressions, and embeds them into the 
calling SQL query. This form is now amenable to cost-based optimization and results 
in efficient, set-oriented, parallel plans as opposed to inefficient, 
iterative, serial execution of UDFs. Froid's approach additionally brings the benefits 
of many compiler optimizations to UDFs with no additional implementation effort. 
We describe the design of Froid and present our experimental evaluation that 
demonstrates performance improvements of up to multiple orders of magnitude on 
real workloads.
\end{abstract}

\section{Introduction} \label{sec:intro}
SQL is arguably one of the key reasons for the popularity of relational 
databases today. SQL's declarative way of expressing intent has on one hand 
provided high-level abstractions for data processing, while on the other 
hand, has enabled the growth of sophisticated query evaluation techniques 
and highly efficient ways to process data. 

Despite the expressive power of declarative SQL, almost all RDBMSs 
support procedural extensions that allow users to write programs in various 
languages (such as Transact-SQL, C\#, Java and R) using imperative 
constructs such as variable assignments, conditional branching, and loops. These 
extensions are quite widely used. For instance, 
we note that there are of the order of tens of millions of Transact-SQL (T-SQL) 
UDFs in use today in the Microsoft Azure SQL Database service, with 
billions of daily invocations.

UDFs and procedures offer many advantages over standard SQL. 
(a) They are an elegant way to achieve modularity and code reuse across SQL queries, 
(b) some computations (such as complex business rules and ML algorithms) are easier to express in 
imperative form, %and 
(c) they allow users to express intent using a mix of simple SQL and imperative 
code, as opposed to complex SQL queries, thereby improving readability and 
maintainability.%, and
%(d) they are used as an access control mechanism, similar to views.
These benefits are not limited to RDBMSs, as
evidenced by the fact that BigData systems (Hive, Spark, etc.) support UDFs as well.

%Database systems support procedural extensions in various languages; however, there has been very 
%little focus towards their efficient evaluation. 
Unfortunately, the above benefits come at a huge performance penalty, 
due to the fact that UDFs are evaluated in a highly inefficient manner.
%as there has been little focus towards efficent evaluation of UDFs.
It is a known fact 
amongst practitioners that UDFs are ``evil'' when it comes to performance 
considerations~\cite{UDFBAD2, RBAR}. In fact, users are advised by experts to 
avoid UDFs for performance reasons. The internet is replete with articles and 
discussions that call out the performance overheads of 
UDFs~\cite{UDFBAD1, UDFBAD3, UDFBAD4, OraBad1, OraBad2}. This is true for all popular RDBMSs, 
commercial and open source.

%On one hand, UDFs are a powerful abstraction, and they encourage good programming 
%practices. On the other hand, poor performance of UDFs 
%due to iterative, interpreted, and serial execution 
%discourages their use. 
UDFs encourage good programming practices and provide a powerful abstraction, and hence
are very attractive to users. But the poor performance of UDFs 
due to na\"{\i}ve execution strategies discourages their use. 
The root cause of poor performance of UDFs can be attributed to
what is known as the `impedance mismatch' between two distinct programming paradigms 
at play -- the declarative paradigm of SQL, 
and the imperative paradigm of procedural code. Reconciling this mismatch is
crucial in order to address this problem, and forms the crux of our paper.

%While programming language compilers perform many optimizations on imperative 
%code, RDBMSs use sophisticated techniques to optimize queries. There are 
%fundamental differences between compilers and query optimizers in their approach 
%to optimization and hence a straightforward application of compiler techniques 
%does not suffice in this situation. Imperative code running in a relational 
%engine brings with it unique challenges and optimization opportunities.

%In this paper, 
We present Froid, an extensible optimization framework for  
imperative code in relational databases. The goal of Froid is to enable developers 
to use the abstractions of UDFs and procedures without 
compromising on performance. Froid achieves this goal using a novel technique to 
automatically convert imperative programs into equivalent relational algebraic 
forms whenever possible. Froid models blocks of imperative code as relational 
expressions, and systematically combines them into a single expression using 
the \textit{Apply}~\cite{Gal01} operator, thereby enabling the query optimizer 
to choose efficient set-oriented, parallel query plans. 

Further, we demonstrate how Froid's relational algebraic transformations can be used to arrive
at the same result as that of applying compiler optimizations (such as dead code 
elimination, program slicing and constant folding) to imperative code. 
Although Froid's current focus is T-SQL UDFs, the 
underlying technique is language-agnostic, and therefore extending it to other 
imperative languages is quite straightforward, as we show in this paper. 

The work of Simhadri et. al~\cite{uudf} was the first to describe 
decorrelation techniques for UDF invocations. Froid improves upon those 
ideas to build a complete optimization framework. 
There have been works that aim to convert fragments of database application
code into SQL in order to improve performance~\cite{Emani2016, Cheung13}.
However, to the best of our knowledge, Froid is the first 
industrial-strength framework that can optimize imperative programs in a relational 
database by transforming them into relational expressions. While Froid is built 
into Microsoft SQL Server, its underlying techniques can be integrated into any RDBMS.
%\footnote{The word 
%`Froid' denotes the rising of a volcano. Our framework is designed as a 
%precursor to a Volcano-based query optimizer and hence the name Froid.}.

We make the following contributions in this paper.
\begin{enumerate}
	\item We describe the unique challenges in optimization of imperative code executing
	in relational databases, and analyze the reasons for their abysmal performance. 
	
	\item We describe the novel techniques underlying Froid, an extensible 
	framework to optimize UDFs in Microsoft SQL Server. We show how Froid integrates 
	with the query processing lifecycle and leverages existing sub-query optimization 
	techniques to transform inefficient, iterative, serial UDF execution strategies into 
	highly efficient, set-oriented, parallel plans.
	
	\item We show how several compiler optimizations such as 
	dead code elimination, dynamic slicing, constant propagation and folding can be 
	expressed as relational algebraic transformations and simplifications that arrive at 
	the same end result. Thereby, Froid brings these additional benefits to
	UDFs with no extra effort. %, by leveraging existing techniques in SQL Server. 
	
	\item We discuss the design and implementation of Froid, and present an %thorough 
	experimental evaluation on several real world customer workloads, showing 
	significant benefits in both performance and resource utilization. 
\end{enumerate}
The rest of the paper is organized as follows. \refsec{sec:bg} gives the background.
Sections~\ref{sec:framework}, \ref{sec:udfalg}, \ref{sec:subst} and \ref{sec:compileropt} describe Froid and its 
techniques. 
%\refsec{sec:compileropt} shows how Froid enables compiler 
%optimizations for UDFs. 
Design details are discussed in
\refsec{sec:impl} followed by an evaluation in~\refsec{sec:eval}. We 
discuss related work in \refsec{sec:relwork} and conclude in~\refsec{sec:concl}.

\section{Background} \label{sec:bg}
In this section, we provide some background regarding 
the way imperative code is currently evaluated in Microsoft SQL Server and analyze the reasons for their poor performance.
%\subsection{Imperative code in SQL Server} \label{subsec:pc}
SQL Server primarily supports imperative code in two forms: UDFs and Stored Procedures (SPs). 
UDFs cannot modify the database state whereas SPs can. 
%Being side-effect free allows UDFs to be used 
%in \textit{select}, \textit{update}, and \textit{delete} statements.
%queries in the SELECT, FROM, WHERE and HAVING clauses.
UDFs and SPs can be implemented in either T-SQL
or Common Language Runtime (CLR). 
T-SQL expands on the SQL standard to include imperative constructs, various 
utility functions, etc. CLR integration allows UDFs and SPs to be written in any 
.NET framework language such as C\#~\cite{CLRU}. 
UDFs can be further classified into two types. Functions that 
return a single value are referred to as scalar  UDFs, and those that 
return a set of rows are referred to as Table Valued Functions (TVFs). SQL Server also 
supports inline TVFs, which are single-statement TVFs analogous to parameterized 
views~\cite{FUNCCREATE}. In this paper we focus primarily on 
\textit{Scalar T-SQL UDFs}. Extensions to support %UDFs and procedures written in 
other imperative languages are discussed in \refsec{sec:extend}.

\subsection{Scalar UDF Example} \label{subsec:example}
\onecolfigure
{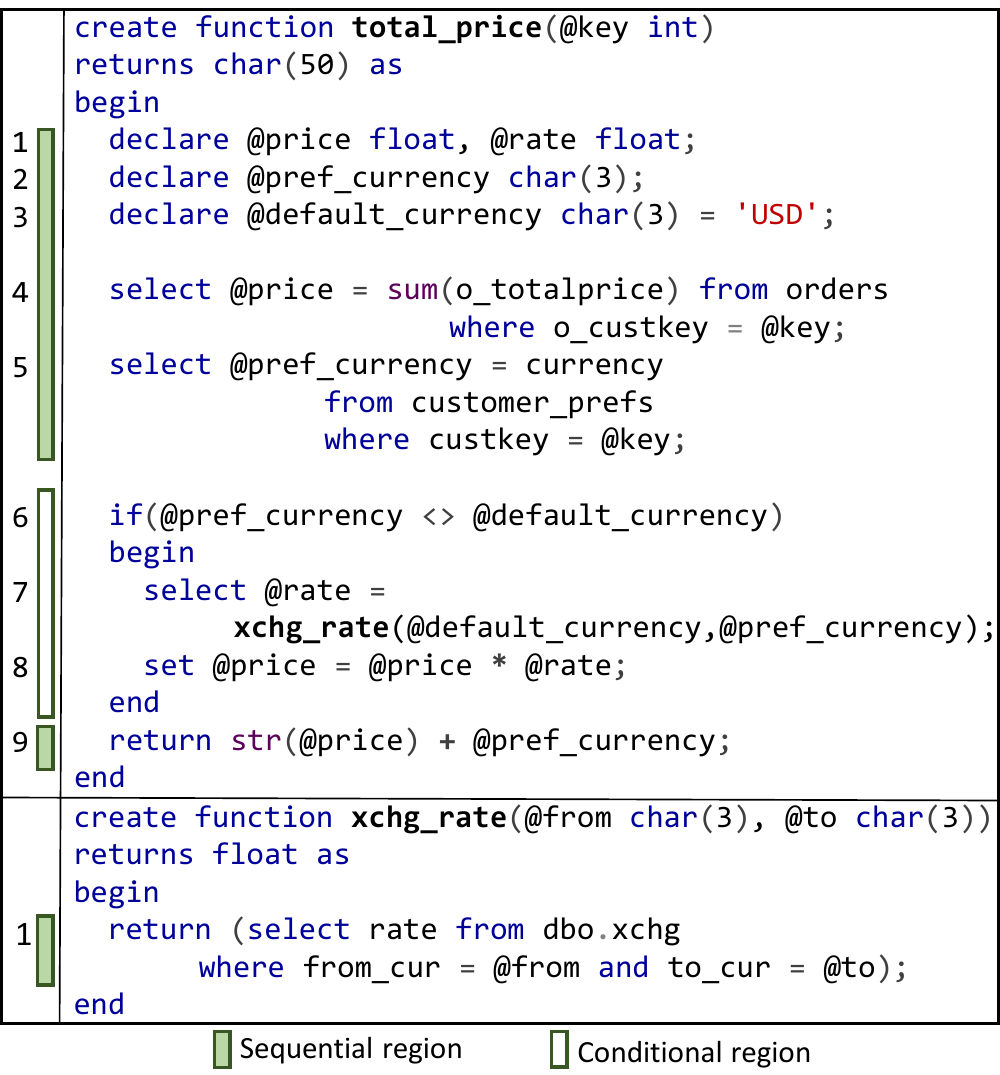}
{Example T-SQL User defined functions}
{fig:udf-example}

In SQL Server, UDFs are created using the CREATE FUNCTION statement \cite{FUNCCREATE}
%. Two such UDFs are 
as shown in \reffig{fig:udf-example}.
The function \textit{total\_price} accepts a customer key, and returns the total 
price of all the orders made by that customer. It computes the price in the 
preferred currency of the customer by looking up the currency code from the 
\textit{customer\_prefs} table and performs currency conversion if necessary.
%In order to retrieve the exchange rate for conversion, 
It calls another UDF 
\textit{xchg\_rate}, that retrieves the exchange rate between the two currencies. 
Finally it converts the price to a string, appends the currency code and returns 
it. Consider a simple query that invokes this UDF.

%Consider a simple query that invokes the UDF \textit{total\_price}.
%in its \textbf{select} 
%clause.

\centerline{\textbf{select} \textit{c\_name, \textbf{dbo.total\_price}(c\_custkey)}}
\centerline{\textbf{from} \textit{customer};}

For each customer, the above query displays the name, and the total price of all 
orders made by that customer. We will use this simple query and the UDFs in \reffig{fig:udf-example}
as an example to illustrate our techniques in this paper.
%From this example, it is easy to see the benefits of UDFs. Although the above
%query can be written by avoiding the UDF, the resulting query 
%(a) would be quite complex and less readable, (b) would not be modular and hence, not reusable, and
%(c) would be hard to modify and hence, less maintainable.

%\section{Evaluation of queries with UDFs} \label{qp}
%\onecolfigurevcrunch
%{figs/basicqp}
%{Query processing pipeline}
%{fig:basicqp}

%In this section, we first describe how UDFs are evaluated in SQL Server.
%Then, we enumerate the reasons for poor performance of UDFs.

\subsection{UDF Evaluation in SQL Server} \label{subsec:udfeval}
We now describe the life cycle of an SQL query that includes a UDF. At 
the outset we note that this is a simplified description with a 
focus on how UDFs are evaluated currently. We refer the reader to 
\cite{conorbook,QPAG,Gal01} for details.
%The key steps in the query processing pipeline are given in \reffig{fig:basicqp}.

\mysubsection{Parsing, Binding and Normalization} \label{qp:pb}
The query first goes through syntactic validation, and is parsed into a tree 
representation. This tree undergoes binding, which includes 
validating referenced objects and loading metadata. Type derivation, view substitution and 
optimizations such as constant folding are also performed. 
Then, the tree is normalized, wherein most common forms of subqueries
are turned into some join variant.
%information about required (implicit) conversions is added at this stage. 
%References to a view are replaced
%with the definition of that view during binding. 
%Some optimizations such as
%constant folding are also performed.
A scalar UDF that appears in a query is parsed and bound as a UDF operator.
The parameters and return type are validated, and metadata 
is loaded. The UDF definition is not analyzed at this stage.

%\mysubsection{Normalization} \label{qp:norm} 
%This step transforms an operator tree into a simplified/normalized form. 
%%Simplifications include, for example, turning outerjoins into joins when 
%%possible, and detecting empty subexpressions. 
%For subqueries, mutual recursion between relational and scalar execution 
%(discussed shortly) is removed, which is always possible; 
%and correlations are removed, which is usually possible. At the end of 
%normalization, most common forms of subqueries are turned into some join variant.

\mysubsection{Cost-based Optimization} \label{qp:qo}
Once the query is parsed and normalized, the query optimizer performs cost-based 
optimization based on cardinality and cost estimates.
%statistics information on indexes and columns. 
%The optimizer estimates the cost for relational operators in the query and 
%associates costs with various alternatives for executing the query. 
Execution alternatives are generated using transformation rules, and the plan with 
the cheapest estimated cost is selected for execution.  
SQL Server's cost-based optimizer follows the design of the 
Volcano optimizer~\cite{GRA93}.
%However, observe that only relational operators are costed and scalar operators
%are not. Prior to the introduction of scalar UDFs, scalars were generally 
%cheap and did not require costing. A small CPU cost added for 
%a scalar operation was enough. This is a cause of bad plan choices only in cases 
%where scalar operations could be arbitrarily expensive, which is often true for 
%scalar UDFs.
%As it explores the plan space, at some point the optimizer might 
%decide that using a current plan is cheaper than continuing to search for 
%something better, and paying the additional cost of continued optimization will 
%not be cost effective. At this point, the best plan found so far is chosen
%for execution.
%Important classes of 
%optimizations include: Reordering of join variants; reordering of GroupBy with 
%join variants; considering special strategies for GroupBy; and introduction of 
%correlated execution (the simplest and most common being index-lookup-join). 
%The architecture of our cost-based optimizer follows the main lines of the 
%Volcano optimizer [9], so that generation of interesting reorderings is done 
%by means of transformation rules.
%\subsubsection*{Plan and procedure cache}
%\noindent\textit{Plan caching:}
%\mysubsection{Plan cache}
SQL Server reuses query plans for queries and UDFs by caching chosen plans. A cache 
entry for a UDF can be thought of as an array of plans, one for each statement in the UDF.

%Since query optimization can be complex and time consuming, SQL Server reuses 
%query plans by caching chosen plans. In addition to ad hoc queries, SQL Server caches plans for 
%UDFs and procedures. A cache entry for a UDF can be
%thought of as an array of plans, one for each statement in the UDF.

\mysubsection{Execution} \label{qp:qe}
The execution engine is responsible for executing the chosen plan efficiently. 
%It is responsible for scheduling threads as necessary, and orchestrating 
%them.
Relational query execution invokes a scalar evaluation sub-system for predicates 
and scalar computations, including scalar UDFs~\cite{GAL07}. 
%The Expression Services library is a component of the execution engine that 
%does data conversion, predicate evaluation (filtering), and arithmetic 
%calculations. Essentially, all scalar expressions including scalar UDFs are 
%evaluated using the expression service. 
The plan for the simple query in~\refsec{subsec:example} is shown in 
\reffig{fig:exampleplan}.
%For a simple query such as the one in our example (\refsec{subsec:example}), 
%the plan typically looks as shown in \reffig{fig:exampleplan}. 
For every tuple that is emitted by the Table Scan operator, the
execution engine calls into the scalar evaluation sub-system to evaluate the
scalar UDF \textit{total\_price}. 

At this point, the execution context switches to the UDF. Now, the UDF
can be thought of as a batch of statements submitted to the engine
%, and each statement is evaluated sequentially. 
If the UDF contains SQL queries (e.g. lines 4 and 5 of \reffig{fig:udf-example}), 
the scalar subsystem makes a recursive call back to the relational execution engine.
Once the current invocation of the UDF completes, the context switches back to the calling query, and the UDF is
invoked for the next tuple -- this process repeats.
%This process repeats for each tuple.
% to the expression 
%service with the result of the UDF. Then, the execution engine calls into
%the expression service again, for the next tuple that came out from the 
%scan operator. This process repeats until the scan has no more rows remaining.
During the first invocation of the UDF, each statement goes through
compilation, and the plan for the UDF 
is cached. During subsequent invocations, 
the cached plan for the UDF is used.

\onecolfigurenogap
{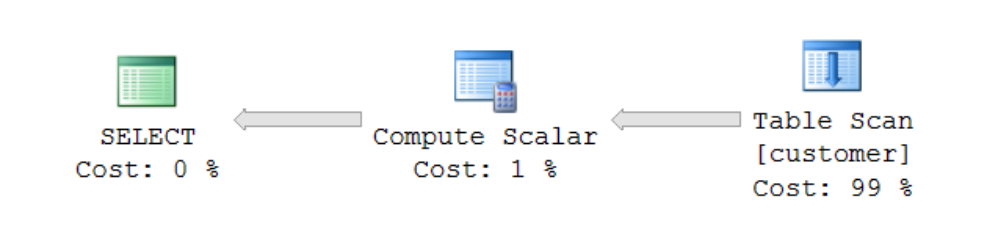}
{Query plan for the query in \refsec{subsec:example}}
{fig:exampleplan}
%\subsection{UDF Performance} \label{sec:udfperf}
\subsection{Drawbacks in UDF Evaluation} \label{sec:udfperf}
%Rename to Performance of Procedural code?
%As mentioned earlier, it is well known that UDFs perform poorly in %most
%relational databases today. 
We now enumerate the main
causes for poor performance of UDFs. % are as follows. %can be enumerated as below.
While we describe the reasons in the context of UDFs in SQL Server, they
are mostly true for other RDBMSs as well, though the finer details may vary.

\mysubsection{Iterative invocation}
UDFs are invoked in an iterative manner, once per qualifying tuple. This
incurs additional costs of repeated context switching due to function 
invocation, and mutual recursion between the scalar evaluation sub-system 
and relational execution. 
Especially, UDFs that execute SQL queries in their
body (which is common in real workloads) are severely affected. 

These iterative plans can be highly inefficient, since queries within the
function body are executed multiple times, once for each invocation.
%These plans can be compared to 
This can be thought of as a nested loops join 
along with expensive context switches and overheads. 
As a consequence, the number of invocations of a UDF in a query has a huge impact
on its performance. The query optimizer is rendered 
helpless here, since it does not look inside UDF definitions.% since UDFs are black boxes to the query optimizer. 
%The impact of iterative execution is more pronounced when
%UDFs are present in the WHERE clause of queries over large tables. 
%In other words, 
%The impact of iterative execution is directly proportional 
%to the number of invocations of the UDF in a query.

\mysubsection{Lack of costing}
Query optimizers treat UDFs as inexpensive black-box operations. 
During optimization, only relational operators are costed, while scalar operators
are not. Prior to the introduction of scalar UDFs, other scalar operators were generally 
cheap and did not require costing. A small CPU cost added for 
a scalar operation was enough. This inadvertent simplification is a crucial cause of 
bad plan choices in cases where scalar operations are arbitrarily expensive, 
which is often true for scalar UDFs.

%Query optimizers treat UDFs as inexpensive black boxes. To 
%the optimizer, the scalar UDF operator is equivalent to any other
%scalar operator (such as an arithmetic operator). Optimizers do not cost 
%scalar UDFs, eventhough they might be arbitrarily expensive. They are 
%in fact, assumed to be cheap, as described in \ref{qp:qo}, which misleads
%the optimizer into choosing bad plans.

\mysubsection{Interpreted execution}
As described in \refsec{qp:qe}, UDFs are evaluated as a batch of statements
that are executed sequentially. In other words, UDFs are 
interpreted statement-by-statement.

Note that each statement itself is compiled, and the compiled plan is 
cached. Although this caching strategy saves some time as it avoids recompilations,
each statement executes in isolation. No cross-statement
optimizations are carried out, unlike in compiled languages.
%As mentioned earlier, 
%There are many optimization techniques for imperative languages that are
%widely used. T
Techniques such as dead code elimination, constant
propagation, folding, etc. have the potential to improve
performance of imperative programs significantly. Na\"{\i}ve evaluation without 
exploiting such techniques is bound to impact performance. 
%All these opportunities are missed by the current approach.

\mysubsection{Limitation on parallelism}
Currently, SQL Server does not use intra-query parallelism in queries 
that invoke UDFs. 
Methods can be designed to mitigate this limitation, but they introduce
additional challenges, such as picking the right degree of parallelism for
each invocation of the UDF.

For instance, consider a UDF that invokes other SQL
queries, such as the one in \reffig{fig:udf-example}. Each such query may itself 
use parallelism,
and therefore, the optimizer has no way of knowing how to share threads
across them, unless it looks into the UDF and decides the degree of 
parallelism for each query within (which could potentially change from one 
invocation to another). With nested and recursive UDFs, this issue becomes even 
more difficult to manage.

\section{The Froid Framework} \label{sec:framework}
As mentioned earlier, Froid is an extensible, language-agnostic optimization
framework for imperative programs in RDBMSs.
The novel techniques behind Froid are able to overcome all the limitations
described above. We now describe the intuition and high level overview of 
Froid. Then, with the help of an example, we walk through the process
of optimizing UDFs in Sections~\ref{sec:udfalg} and~\ref{sec:subst}.

\subsection{Intuition}
Queries that invoke UDFs, such as the one in \refsec{subsec:example} can be thought of as
queries with complex sub-queries. In nested sub-queries, the inner
query is just another SQL query (with or without correlation). UDFs
on the other hand, use a mix of imperative language constructs and SQL, and 
hence are more complex. The key observation here is that 
iterative execution of UDFs is similar to correlated evaluation of nested 
sub-queries. Froid is based on this observation which was made in~\cite{GUR09} and~\cite{uudf}.

Optimization of sub-queries has received a lot of attention 
in the database literature and industry (see \refsec{sec:relwork} for details). 
In fact, many of the popular RDBMSs are able to 
transform correlated sub-queries into joins, thereby enabling the 
choice of set-oriented plans instead of iterative evaluation of sub-queries. 

Given these observations, the intuition behind Froid can be succintly stated as follows.
\textit{If the entire body of an imperative UDF can be expressed as a 
single relational expression R, then any query that invokes this UDF can be 
transformed into a query with R as a nested sub-query in place of the UDF.} 
We term this semantics-preserving transformation as \textit{unnesting} or 
\textit{inlining} of the UDF into the calling query. 

Once we perform this transformation, we can leverage existing sub-query 
optimization techniques to get better plans for queries with UDFs. This 
transformation forms the crux of Froid. Note that although we use the term 
\textit{inlining} to denote this transformation, it is fundamentally 
different compared to inlining in imperative programming languages.

\subsection{The APPLY operator}
Froid makes use of the \textit{Apply} operator while building a relational 
expression for UDFs. Specifically, it is used to
combine multiple relational expressions into a single expression.
The \textit{Apply} operator ($\apply$) was originally designed to model correlated execution of 
sub-queries algebraically in SQL Server~\cite{Gal01, GAL07}. It accepts a 
relational input \textit{R} and a parameterized relational 
expression $E(r)$. For each row $r \in R$, it evaluates $E(r)$ and emits tuples as a join 
between $r$ and $E(r)$. More formally, it is defined as follows~\cite{Gal01}: 
%The formal definition of the {\em Apply} operator ($\apply$), as 
%defined in~\cite{Gal01} is as follows: 
\[ R\ \apply^{\otimes}\ E = \bigcup_{r \in R} (\{r\} \otimes E(r)) \]
where $\otimes$, known as the join type, is either cross product, left outer-join, 
left semijoin or left antijoin.
SQL Server's query optimizer has a suite of transformation rules for sub-query 
decorrelation, which remove the \textit{Apply} operator and enable the use of 
set-oriented relational operations whenever possible. Details with examples can be found 
in~\cite{Gal01,GAL07,uudf}.
%In prior work~\cite{uudf}, the authors define extensions 
%to the \textit{Apply} operator to model UDFs algebraically. 
%Froid does not require any new operator extensions. For a detailed comparison 
%with~\cite{uudf}, we refer the reader to \refsec{sec:relwork}.

\subsection{Overview of Approach} \label{subsec:overview}
\onecolfigure%vcrunch
{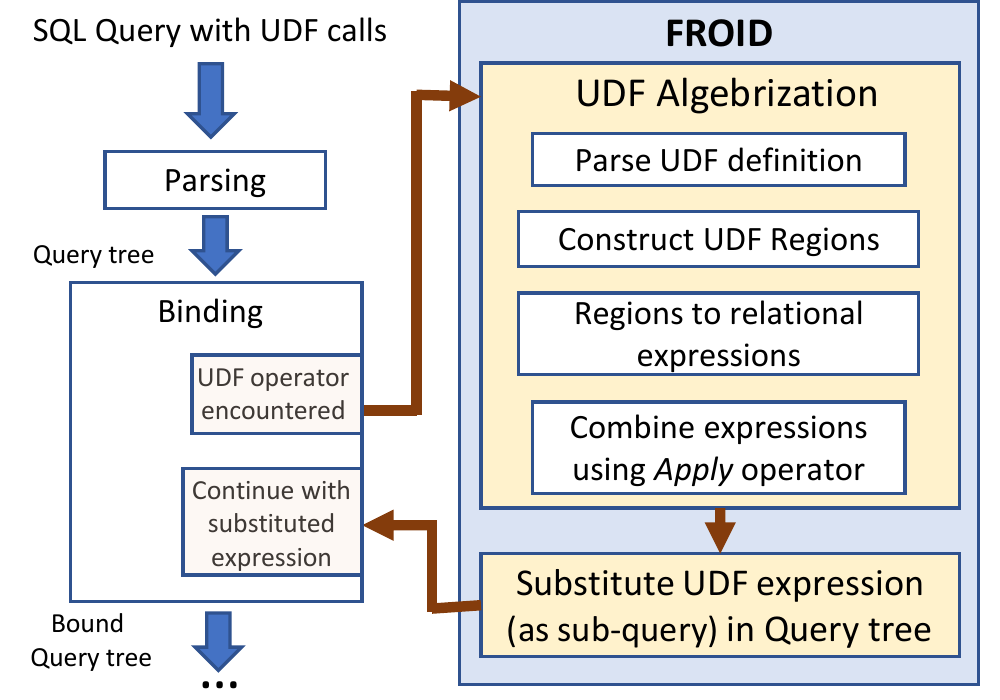}
{Overview of the Froid framework}
{fig:flare-arch}

For a UDF with a single RETURN statement in its body, such as the function \textit{xchg\_rate}
in \reffig{fig:udf-example}, the transformation is straightforward. The body of such a 
UDF is already a single relational expression, and therefore it can be substituted
easily into the calling context, like view substitution.% (similar to the substitution of views and inline TVFs).

Expressing the body of a multi-statement UDF (such as the function \textit{total\_price} 
in \reffig{fig:udf-example}) as a single relational expression is a non-trivial task. 
Multi-statement UDFs typically use imperative constructs such as variable 
declarations, assignments, conditional branching, and loops. 
Froid models individual imperative constructs as relational 
expressions and systematically combines them to form one expression. 

%\mysubsection{Transformation phases}
\reffig{fig:flare-arch} 
depicts the high-level approach of Froid, consisting of two phases: 
UDF algebrization followed by substitution.
As a part of binding, the query tree is traversed and each node is bound, as 
described in \refsec{qp:pb}. During binding, if a UDF operator is encountered, 
the control is transferred to Froid, and UDF algebrization is initiated. UDF 
algebrization involves parsing the statements of the UDF and constructing an 
equivalent relational expression for the entire UDF body (described in \refsec{sec:udfalg}). 
This resulting expression is then substituted, or embedded in the query tree of 
the calling query in place of the UDF operator (described in \refsec{sec:subst}). 
This query tree with the substituted UDF expression is bound using
the regular binding process. If references to other (nested) 
UDF operators are encountered, the same process is repeated. This transformation finally results 
in a bound query tree, which forms the input to normalization and optimization. 

\renewcommand{\arraystretch}{1.2}

	\begin{table*}[ht]	
		\centering
		\small
		\begin{tabular}{|@{ }c@{ }|@{ }c@{ }|}\hline
			\textbf{Imperative Statement (T-SQL)}& \textbf{Relational expression (T-SQL)} \\
			\hline 
			$ \text{DECLARE}\ \{@var\ data\_type\, [= expr]\} [,\ldots n];$ & $ \text{SELECT}\ \{expr|null\ \text{AS}\ var\} [,\ldots n];$ \\%[.1em]
			\hline 
			$ \text{SET}\ \{@var = expr\} [,\ldots n]; $ & $ \text{SELECT}\ \{expr\ \text{AS}\ var\} [,\ldots n];$ \\%[.1em]
			\hline
			$ \text{SELECT}\ \{@var1 = prj\_expr1\} [,\ldots n]\ \text{FROM}\ sql\_expr;$ & $ \{\text{SELECT}\ prj\_expr1\ \text{AS}\ var1\ \text{FROM}\ sql\_expr\};\ [,\ldots n] $ \\%[.1em]
			\hline
			\multirow{2}{*}{$ \text{IF } (pred\_expr)\quad   \{t\_stmt;[\ldots n]\}$ $\text{ELSE}\quad    \{f\_stmt;[,\ldots n]\}$} & $\text{SELECT CASE WHEN}\ pred\_expr \text{ THEN 1 ELSE 0 END AS}\  pred\_val;$ \\ & \{$\text{SELECT CASE WHEN }pred\_val = 1 \text{ THEN } t\_stmt \text{ ELSE } f\_stmt;\}[\ldots n] $ \\%[.1em]
			\hline
			$ RETURN\ expr;$ & $ \text{SELECT}\ expr\ \text{AS}\ returnVal;$ \\%[.1em]
			\hline
		\end{tabular}
		\caption{Relational algebraic expressions for imperative statements (using standard T-SQL notation from \protect\cite{TSQL})}
		\label{tab:imprel}
		\vspace{-1.5em}
	\end{table*}

\renewcommand{\arraystretch}{1.0}

\subsection{Supported UDFs and queries} \label{subsec:support}
%\mysubsection{Language constructs}
Froid currently supports the following imperative constructs in scalar UDFs.
%that can address a fairly large class of UDFs encountered in practice.
%Froid currently supports the following imperative constructs:
%\begin{itemize}[leftmargin=*,noitemsep]
%	\item Variable declaration and assignments.
%	\begin{align*}
%		DECLARE\ \{&@var\ data\_type\, [= expr]\} [,\ldots n];\\
%		SET\ \{&@var = expr\} [,\ldots n]; 
%	\end{align*}
%	\item SQL query, with multiple variable assignments.
%	\begin{align*}
%		SELECT\ {@var1 = sql\_expr} [,\ldots n];
%	\end{align*}
%	\item Conditional branching with arbitrary levels of nesting.
%	\begin{align*}
%		IF\ (pred\_expr)\quad   \{stmt;[\ldots n]\}\quad  ELSE\quad    \{stmt;[,\ldots n]\}			
%	\end{align*}	
%	\item Single or multiple return statements.
%	\begin{align*}
%		RETURN\ expr;
%	\end{align*}
%	\item Nested/recursive UDF invocations.
%	\item Other relational operations such as EXISTS, ISNULL.	
%\end{itemize}

\begin{itemize}[leftmargin=*,noitemsep]
	\item \textbf{DECLARE, SET:} Variable declaration and assignments.
	\item \textbf{SELECT:} SQL query with multiple variable assignments.
	\item \textbf{IF/ELSE:} Branching with arbitrary levels of nesting.
	\item \textbf{RETURN:} Single or multiple return statements.
	\item \textbf{UDF:} Nested/recursive function calls.
	\item \textbf{Others:} Relational operations such as EXISTS, ISNULL.	
\end{itemize}

%UDFs can contain cursor loops that iterate over query results, or arbitrary loops.
%Apart from simple cursor loops, it is hard to express loops as relational
%expressions~\cite{uudf}. 
Table \ref{tab:imprel} (column 1) shows the supported constructs more formally.
In Table \ref{tab:imprel}, $@var$ and $@var1$ denote variable names, $expr$ is any valid T-SQL expression including
a scalar subquery; $prj\_expr$ represents a projected column/expression; 
$sql\_expr$ is any SQL query; $pred\_expr$ is a boolean expression; $t\_stmt$ and $f\_stmt$
are T-SQL statements~\cite{TSQL}.

%\mysubsection{UDF and Query complexity}
Froid's techniques do not impose any limitations on the size or depths of UDFs and complexity 
of queries that invoke them. The only precondition for our transformations is that the UDF 
has to use the supported constructs. However, in practice, there are certain special cases 
where we partially restrict the application of our transformations; they are discussed 
in~\refsec{subsec:limits}.

\section{UDF Algebrization} \label{sec:udfalg}
We now describe the first phase of Froid in detail. The goal here
is to build a single relational expression which is 
semantically equivalent to the UDF. This involves transforming 
imperative constructs into equivalent relational expressions and combining 
them in a way that strictly adheres to the procedural intent of the UDF. 
UDF algebrization consists of the following three steps.

\subsection{Construction of Regions} \label{subsec:region}
First, each statement in the UDF is parsed and the body of the UDF is divided
into a hierarchy of program \textit{regions}. Regions represent structured fragments of 
programs such as basic blocks, if-else blocks and loops~\cite{Hecht72}. Basic blocks
are referred to as sequential regions, if-else blocks are referred to as 
conditional regions, and loops are referred to as loop regions. Regions by 
definition contain other regions; the UDF as a whole is also a region. 

Function \textit{total\_price} of \reffig{fig:udf-example} is a sequential 
region R0 (lines 1-9). It is in turn composed of three consecutive sub-regions 
denoted R1, R2 and R3. R1 is a sequential region (lines 1-5), R2 is a conditional 
region (lines 6-8), and R3 is a sequential region (line 9) as indicated in \reffig{fig:udf-example}. 
Regions can be constructed in a single pass over the UDF body.

\subsection{Relational Expressions for Regions}
Once regions are constructed, the next step is to construct a relational 
expression for each region. 

%In order to strictly adhere to the procedural semantics 
%of the UDF, Froid has to ensure that any computation in the relational 
%equivalent of the UDF occurs if and only if that computation would have occured 
%in the imperative version of the UDF. 
%We first consider 
%individual imperative statements and show how Froid constructs relational 
%expressions for them.

\subsubsection{Imperative statements to relational expressions} \label{subsubsec:imp2rel}
%We now show how Froid first constructs relational expressions for individual 
%imperative statements, and then combines them to form an expression for
%a region. 
%The equivalence of these individual transformations follow from 
%the semantics of operations involved.
Froid first constructs relational expressions for individual 
imperative statements, and then combines them to form a single expression for a region.
These constructions make use of the \textit{ConstantScan} 
and \textit{ComputeScalar} operators in SQL Server~\cite{LPOR}. 
The \textit{Constant\-Scan} operator introduces one row with no column. A 
\textit{Compu\-te\-Scalar}, typically used after a \textit{ConstantScan}, adds 
computed columns to the row. 
%\footnote{There are similar operators available in other databases}.

\vspace{0.3mm}
\mysubsection{Variable declarations and assignments}
The T-SQL constructs DECLARE, SET and SELECT fall under this category. These 
statements are converted into relational equivalents by modeling them as 
projections of computed columns in relational algebra as shown in Table~\ref{tab:imprel} (rows 1, 2, 3).
%In SQL Server, this is accomplished by using a \textit{ConstantScan} operator 
%followed by a \textit{ComputeScalar}. The \textit{ComputeScalar} operator 
%captures the scalar expression on the RHS of the assignment. 
For example, consider line 3 of \reffig{fig:udf-example}: \\
\centerline{\textbf{set }\textit{@default\_currency} = \textit{`USD'};}

\noindent This is represented in relational form as\\
\centerline{\textbf{select }\textit{`USD'} \textbf{as} \textit{default\_currency}.}

Observe that program variables are transformed into attributes projected by the 
relational expression. The RHS of the assignment could be any scalar expression 
including a scalar valued SQL query (when the SELECT construct is used). In this 
case, we construct a \textit{ScalarSubQuery} instead of \textit{ComputeScalar}. 
For example, the assignment statement in line 4 of \reffig{fig:udf-example} is 
represented in relational form as\\
\centerline{\textbf{select}\textit{(}\textbf{select} \textit{sum(o\_totalprice)} \textbf{from} \textit{orders}} \\
\centerline{\textbf{where }\textit{o\_custkey = @key)} \textbf{as} \textit{price}}.

Variable declarations without initial assignments are considered as 
assignments to \textit{null} or the default values of the corresponding data types.
Note that the DECLARE and SELECT constructs can assign to one or more variables in 
a single statement, but Froid handles them as multiple assignment statements. 
Modeling them as multiple assignment statements might lead to RHS expressions
being repeated. However, common sub-expression elimination can remove such duplication in most cases.

\mysubsection{Conditional statements}
A conditional statement is typically specified using the IF-ELSE T-SQL construct.
It consists of a predicate, a \textit{true} block, and a \textit{false} block. 
This can be algebrized using SQL Server's CASE construct as given in Table~\ref{tab:imprel} (row 4). 
The \textit{switch-case} imperative construct is also internally expressed as the
IF-ELSE construct, and behaves similarly. Consider the following example:\\
\phantom{1tab}\phantom{1tab1tab}\textbf{if}\textit{(@total $>$ 1000)}\\
\phantom{1tab}\phantom{1tab1tab1tab}\textbf{set }\textit{@val = `high'; }\\
\phantom{1tab}\phantom{1tab1tab}\textbf{else} \\%\textbf{set }\textit{@val = `low'; }
\phantom{1tab}\phantom{1tab1tab1tab}\textbf{set }\textit{@val = `low'; }
%\centerline{\textit{if(@total $>$ 1000) set @val = `high'; }}
%\centerline{\textit{else set @val = `low';}}
%\begin{Verbatim}[obeytabs, tabsize=2]
%				if(@total > 1000) 
%				  set val = `high'; 
%				else 
%				  set val = `low'; 
%\end{Verbatim}

\noindent The above statement is represented in relational form as \\
\centerline{\textbf{select}\textit{(}\textbf{case when } \textit{total $>$ 1000} \textbf{then } \textit{`high'}}
\centerline{\textbf{else }\textit{`low'} \textbf{ end} \textit{)} \textbf{as }\textit{val}.}
%\centerline{\textbf{else }\textit{`low'} \textbf{ end} \textit{)} \textbf{as }\textit{val}.}

This approach works for simple cases. For complex and nested conditional blocks,
this approach may lead to redundant 
computations of the predicate thereby violating the procedural intent of the UDF. 
Re-evaluating a predicate multiple times not only goes against our principle
of adherence to intent, but it might also hurt performance if the predicate
is expensive to evaluate. 
Froid addresses this by assigning the value of the
predicate evaluation to an implicit boolean variable (shown as $pred\_val$ in row 4 of Table~\ref{tab:imprel}). Subsequently, whenever 
necessary, it uses the CASE expression to check the value of this implicit 
boolean variable. %This is similar to the \textit{if-conversion} technique in compilers.

\mysubsection{Return statements}
Return statements denote the end of function
execution and provide the value that needs to be returned from the
function. Note that a UDF may have multiple return statements, one per
code path.
Froid models return statements as assignments to an implicit variable called 
\textit{returnVal} (shown in row 5 of Table~\ref{tab:imprel}) followed by an unconditional jump to the end of the UDF. 
%The return type of the UDF forms the data type of this variable. 
This 
unconditional jump means that no statement should be evaluated once the
\textit{returnVal} has been assigned a valid return value (note that \textit{null}
could also be a valid return value). 
Froid implicitly declares the variable \textit{returnVal} at the first occurrance 
of a return statement. Any subsequent occurrance of a return statement is treated as an 
assignment to \textit{returnVal}. 

Froid models unconditional jumps using the \textit{probe} and \textit{pass-through}
functionality of the \textit{Apply} operator~\cite{GAL07}. The \textit{probe} is used
to denote whether \textit{returnVal} has been assigned, and the
\textit{pass-through} predicate ensures that subsequent operations are
executed only if it has not yet been assigned.

Although unconditional jumps could be modeled without
using \textit{probe} and \textit{pass-through}, there are disadvantages to that
approach.
%Note that this behavior could be expressed using \textit{case} expressions without
%using \textit{probe} and \textit{pass-through}. However, there are two disadvantages 
%of using \textit{case} expressions to implement unconditional jumps. 
First, it increases 
the size and complexity of the resulting expression.
%, the resulting expression could end up very huge and complex, 
%depending upon the number of return statements in the UDF. 
This is because all successor regions of a return 
statement would need to be wrapped within a \textit{case} expression.
Second, the introduction of \textit{case} expressions hinders
the applicability of scalar expression folding and simplification. As we shall describe in 
\refsec{sec:compileropt}, Froid brings optimizations such as constant folding and constant 
propagation to UDFs. The applicability of these optimizations would be
restricted by the use of \textit{case} expressions to model unconditional jumps. 
%We omit an example due to lack of space.

\mysubsection{Function invocations}
Functions may invoke other functions, and may be recursive as well. Froid can unnest such 
nested function calls to achieve more gains. When a function invocation statement is 
encountered during UDF algebrization, Froid simply retains the UDF operator as the 
relational expression for that function. As part of the normal binding process in SQL 
Server, Froid is again invoked for the nested function, thereby inlining it as well. Some special cases 
with deeply nested/recursive functions, where we choose not to optimize are discussed 
in \refsec{subsec:limits}. 

%\mysubsection{Loops}
%UDFs can contain cursor loops that iterate over query results, or arbitrary loops.
%Apart from simple cursor loops, it is hard to express loops as relational
%expressions~\cite{uudf}. We have prototyped loop algebrization in Froid 
%for simple cursor loops. However, from our analysis of many real world workloads,
%we found that scalar UDFs with loops are quite rare (see~\refsec{sec:eval}).
%Therefore, we have currently disabled support for loops and may enable it in future.

\mysubsection{Others}
Relational operations such as EXISTS, NOT EXISTS, ISNULL etc.
can appear in imperative constructs such as the predicate of an IF-ELSE
block. Froid simply uses the corresponding relational operators in these 
cases.
In addition to the above constructs, we have prototyped algebrization of  
cursor loops. However, from our analysis of many real world workloads,
we found that scalar UDFs with loops are quite rare (see~\refsec{sec:eval}).
Therefore, we have currently disabled support for loops and may enable it in future.

\subsubsection{Derived table representation} \label{subsub:dt}
\begin{table}
	\small
	\centering	
	\begin{tabular}{ |c|p{6.5cm}| } 
		\hline
		\textbf{Region} & \textbf{Write-sets (Derived table schema)} \\
		\hline
		R1 & DT1 (price \textit{float}, rate \textit{float}, \newline default\_currency \textit{char(3)}, pref\_currency \textit{char(3))} \\
		\hline
		R2 & DT2 (price \textit{float}, rate \textit{float}) \\
		\hline
		R3 & DT3 (returnVal \textit{char(50)}) \\
		\hline
	\end{tabular}
	\caption{Derived tables for regions in function \textit{total\_price}.}	
	\label{tab:dt} 
	\vspace{-1.5em}
\end{table}

We now show how expressions for individual statements are combined into
a single expression for a region using derived tables.
A \textit{derived table} is a statement-local temporary table created by 
a subquery. Derived tables can be aliased and referenced just like normal tables. 
Froid constructs the expression of each region as a derived table
as follows.

Every statement in an imperative program has a `Read-Set' and a `Write-Set', 
representing sets of variables that are read from and written to within that statement 
respectively. Similarly, every region R can be seen as a compound statement 
that has a Read-Set and a Write-Set. Informally, the Read-Set of region R is the union of the 
Read-Sets of all statements within R. The Write-Set of R is the union of the
Write-Sets of all statements within R.

A relational expression that captures the semantics of a region R has to expose 
the Write-Set of R to its subsequent regions. This is because the variables 
written to in region R would be read/modified in subsequent regions of the UDF. 
The Write-Set of region R is therefore used to define the schema of the relational
expression for R. The schema is defined by treating every variable in the 
Write-Set of R as an attribute. The implicit variable \textit{returnVal}
appears in the Write-Set of all regions that have a RETURN statement.

The Write-Sets of all the regions  in function \textit{total\_price} of 
\reffig{fig:udf-example} are given in Table~\ref{tab:dt}.
Using the schema, along with the relational expressions for each 
statement, we can construct a relational expression for the entire
region R. A single \textit{ConstantScan} followed by 
\textit{ComputeScalar} operators, one per variable, results in a derived
table with a single tuple. This derived table represents the values of 
all variables written to in R. The derived table aliases for regions
R1, R2 and R3 are shown as DT1, DT2, and DT3 in Table~\ref{tab:dt}.

\subsection{Combining expressions using APPLY} 
\onecolfigure
{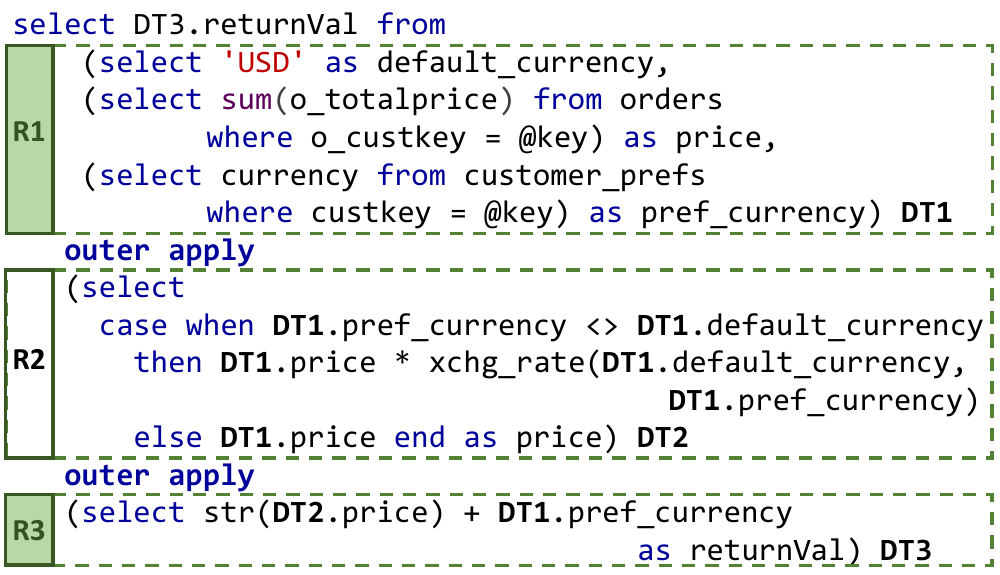}
{Relational expression for UDF total\_price}
{fig:apply-tree}

Once we have a relational expression per region, we now proceed to
create a single expression for the entire function. The relational expression 
for a region R uses attributes from its prior regions, and exposes its attributes to 
subsequent regions. Therefore, we need a mechanism to connect variable 
definitions to their uses and (re-)definitions. %\footnote{A write to a variable is called a `definition', and a read of its value is called a `use' in compiler terminology.}. 

Froid makes use of the relational \textit{Apply} operator to systematically 
combine region expressions. The derived tables of each region 
are combined depending upon the type of the parent region. 
For a region $R$, we denote the corresponding relational expression as $E(R)$.
For the \textit{total\_price} function in \reffig{fig:udf-example}, 
$E(R1) = DT1, E(R2) = DT2, E(R3) = DT3$.

\reffig{fig:apply-tree} shows the relational expression for 
the entire UDF. The dashed boxes in 
\reffig{fig:apply-tree} indicate relational expressions for individual
regions R1, R2 and R3.
Note that Froid's transformations are performed on the relational query tree
structure and not at the SQL language layer. 
\reffig{fig:apply-tree} shows an SQL representation for ease of presentation.

The relational expression for a sequential region such as R0 is constructed
using a sequence of \textit{Apply} operators between its consecutive sub-regions i.e.,
\[ E(R0) = (E(R1)\  \mathcal{A}^o\ E(R2))\ \mathcal{A}^o\ E(R3) \]
The SQL form of this equation can be seen in \reffig{fig:apply-tree}. The 
\textit{Apply} operators make the values in DT1 
available for use in DT2, the values in DT1 and DT2 available for DT3, and so 
on. We use the outer join type for these \textit{Apply} operators ($\mathcal{A}^o$). In the 
presence of multiple return statements, we make use of \textit{Apply} with 
\textit{probe} (which internally uses left semijoin) and \textit{pass-through} 
(outer join)~\cite{GAL07}.

Consider the variable \textit{@pref\_currency} as an example. It is first 
computed in R1, and hence is an attribute of the derived table DT1 (as shown in 
\reffig{fig:apply-tree}). R2 uses this variable, but does not modify it. Therefore 
\textit{@pref\_currency} is not in the schema of DT2. All the uses of 
\textit{@pref\_currency} in R2 now refer to it as \textit{DT1.pref\_currency}. 
R3 also uses \textit{@pref\_currency} but does not modify it. The value of 
\textit{@pref\_\-currency} that R3 uses comes from R1. Therefore R3 also makes 
use of \textit{DT1.pref\_currency} in its computation of \textit{returnVal}.

Observe that the expression in \reffig{fig:apply-tree} has no reference to 
the intermediate variable \textit{@rate}. As a simplification, we generate
expressions for variables only when they are first assigned a value, and 
we expose only those variables that are live at the end of the region (i.e., used subsequently). 
The \textit{@rate} variable gets eliminated due to these simplifications.
Finally, observe that the only attribute exposed by R0 (the entire function) is 
the \textit{returnVal} attribute. This expression shown in \reffig{fig:apply-tree}, is
a relational expression that returns a value equal to the return value 
of the function \textit{total\_price}.

\subsection{Correctness and Semantics Preservation} \label{sec:semantics}
We now reason about the correctness of our transformations, and describe how 
they preserve the procedural semantics of UDFs. As described earlier, Froid 
first constructs equivalent relational expressions for individual imperative 
statements (\refsec{subsubsec:imp2rel}). The correctness of these individual 
transformations directly follows from the semantics of the imperative 
construct being modeled, and the definition of the relational operations used 
to model it. The updated values of variables due to assignments
are captured using derived tables consisting of a single tuple of values.

Once individual statements (and regions) are modeled as single-tuple relations
(\refsec{subsub:dt}), performing an \textit{Apply} operation
between these relations results in a single-tuple relation, by definition.
By defining derived table aliases for these single-tuple relations and using the
appropriate aliases, we ensure that all the data dependencies 
are preserved. The relational \textit{Apply} operator is composable, 
allowing us to build up more complex expressions using previously built 
expressions, while maintaining correctness.

In order to strictly adhere to the procedural intent of the UDF, Froid 
ensures that any computation in the relational equivalent of the UDF occurs only 
if that computation would have occurred in the procedural version of the UDF. This
is achieved by (a) using the \textit{probe} and \textit{pass-through} extensions 
of the \textit{Apply} operator to ensure that unconditional jumps are respected, 
(b) avoiding re-evaluation of predicates by assigning their results into implicit 
variables, and (c) using CASE expressions to model conditional statements.
%instead of WHERE clauses.

\section{Substitution and Optimization} \label{sec:subst}
Once we build a single expression for a UDF, the high-level approach to embed 
this expression into the calling query is similar to view substitution, typically 
done during binding. Froid replaces the 
scalar UDF operator in the calling query with the newly constructed relational 
expression as a scalar sub-query. The parameters of the UDF (if any) form the 
correlating parameters for the scalar sub-query. At substitution time, references 
to formal parameters in the function are replaced by actual parameters from the 
calling query. 
SQL Server has sophisticated optimization techniques for subqueries~\cite{Gal01}, 
which can be then leveraged. 
In fact, SQL Server never chooses to do correlated 
evaluation for scalar valued sub-queries~\cite{GAL07}. 
The plan (with Froid enabled) for the query 
in \refsec{subsec:example} is given in \reffig{fig:inlined-plan}.
Although this plan is quite complex compared to the simple plan in \reffig{fig:exampleplan},
it is significantly better.
From the plan, we observe that the optimizer has (a) inferred the 
joins between \textit{customer}, \textit{orders}, \textit{customer\_prefs} and \textit{xchg} -- all of which 
were implicit, (b) inferred the appropriate \textit{group by} operations
and (c) parallelized the entire plan.

Froid overcomes all the four limitations in UDF evaluation enumerated in \refsec{sec:udfperf}.
%As described in \refsec{sec:udfperf}, there are four key reasons for poor performance of UDFs. Froid overcomes all the four limitations. 
First, the optimizer now decorrelates the scalar sub-query and chooses set-oriented plans 
avoiding iterative execution. Second, expensive operations inside the UDF 
are now visible to the optimizer, and are hence costed. Third, the UDF is no 
longer interpreted since it is now a single relational expression. Fourth, the 
limitation on parallelism no longer holds since the entire query including the 
UDF is now in the same execution context.

In %an industrial strength 
a commercial database with a large user base such as 
SQL Server, making intrusive changes to the query optimizer can have unexpected 
repercussions and can be extremely risky. 
One of the key advantages of Froid's approach is that it requires no changes
to the query optimizer. It leverages existing query optimization rules
and techniques by transforming the imperative program into a form that the 
query optimizer already understands.

\twocolfigure
{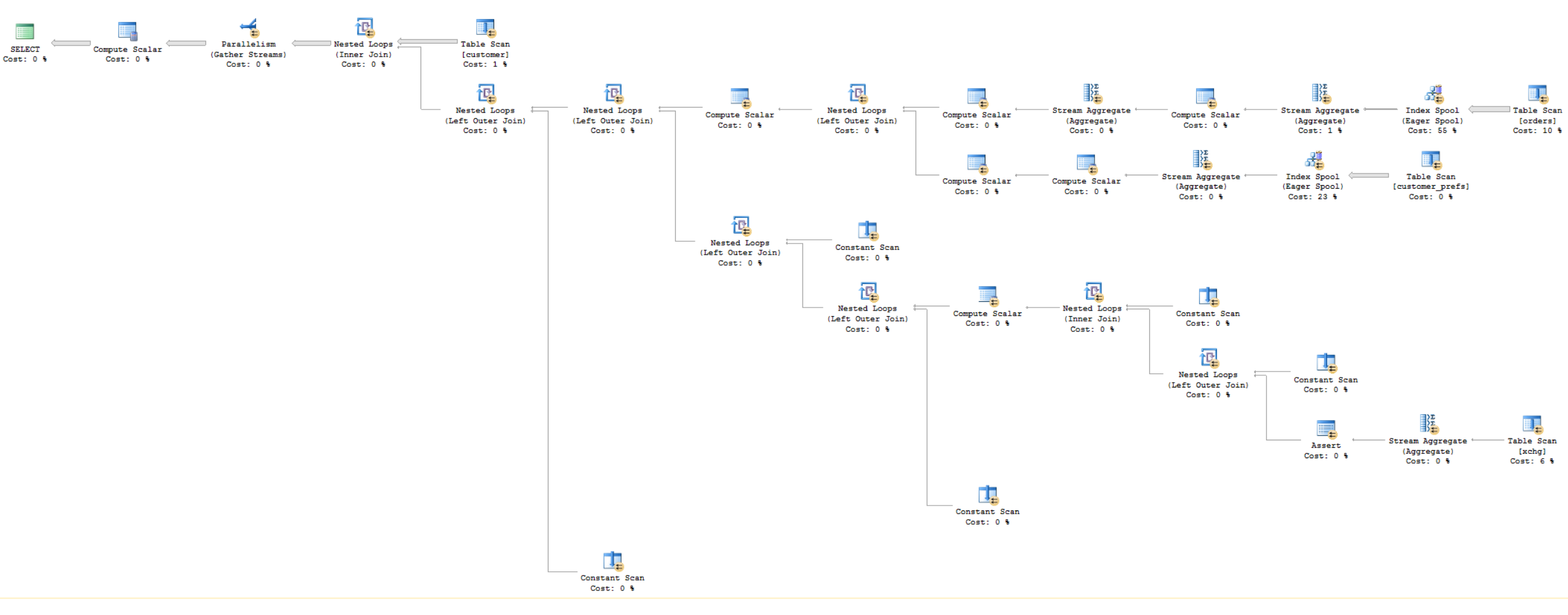}
{Plan for inlined UDF total\_price of Figure 1}
{fig:inlined-plan}

\section{Compiler Optimizations} \label{sec:compileropt}
\twocolfigure
{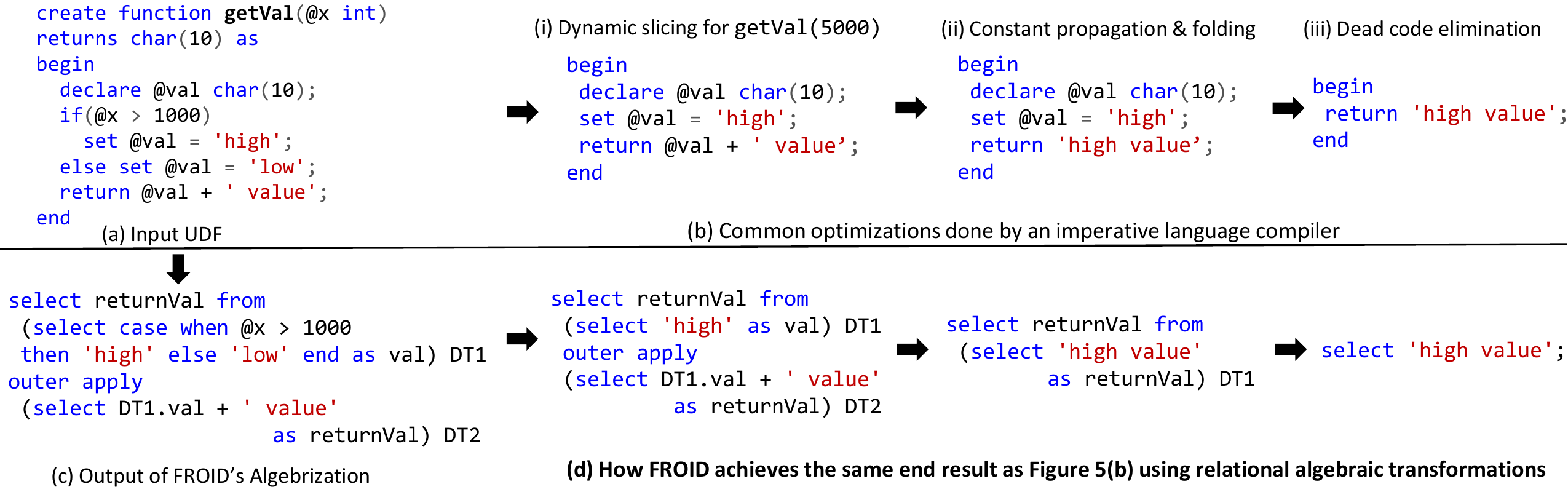}
{Compiler optimizations as relational transformations. For ease of presentation, (c) and (d) are shown in SQL; these are actually transformations on the relational query tree representation.}
{fig:slicing}

Froid's approach not only overcomes current 
drawbacks in UDF evaluation, but also adds a bonus: \textit{with no additional implementation effort, 
it brings to UDFs the benefits of several optimizations done by an imperative language compiler}.
In this section, we point out how some common optimization techniques 
for imperative code can be expressed as relational algebraic transformations
and simplifications. Due to this, Froid is able to achieve these additional 
benefits by leveraging existing sophisticated query optimization techniques 
present in Microsoft SQL Server. 

Using a simple example, \reffig{fig:slicing} illustrates the working of Froid's transformations in contrast with 
compiler optimizations. The function \textit{getVal} (\reffig{fig:slicing}(a)) 
sets the value of variable \textit{@val} based on a predicate. Starting with this UDF, a few common optimizations 
done by an imperative language compiler are shown in \reffig{fig:slicing}(b) in three steps. 
Starting from the same input UDF, \reffig{fig:slicing}(c) shows the output
of Froid's algebrization. Then, \reffig{fig:slicing}(d) shows 
relational algebraic transformations such as projection-push\-down and apply-removal
that Froid uses, to arrive at the same result as the compiler optimizations in \reffig{fig:slicing}(b).

\subsection{Dynamic Slicing}
Dynamic slicing is a program slicing technique that makes use of information 
about a particular execution of a program. A dynamic slice for a program contains 
a subset of program statements that will be visited in a particular execution of 
the program~\cite{KENNEDYBOOK,MUCHNICK}.
For a particular invocation of the UDF in \reffig{fig:slicing}(a), only one of 
its conditional branches is taken. For example, the dynamic slice for 
\textit{getVal(5000)} is given in \reffig{fig:slicing}(b)(i). As we can observe 
from \reffig{fig:slicing}(d), Froid achieves slicing by evaluating the predicate 
($@x > 1000$) at compile time and removing the case expression. In such cases 
where one or more parameters to a UDF are compile time constants, Froid simplifies 
the expression to use the relevant slice of the UDF by using techniques such as 
projection pushdown and scalar expression simplification. 

\subsection{Constant Folding and Propagation}
Constant folding and constant propagation are related optimizations used by modern 
compilers~\cite{Aho06, KENNEDYBOOK}. Constant folding is the process of recognizing 
and evaluating constant expressions at compile time. Constant propagation is the 
process of substituting the values of known constants in expressions at compile 
time. 
%Constant folding and propagation are typically used together to achieve 
%many simplifications and reductions, by interleaving them iteratively.

SQL Server already performs constant folding within the scope of a single statement. 
However, since it does not perform cross-statement optimizations, constant 
propagation is not possible. This leads to re-evaluation of many 
expressions for every invocation of the UDF. Froid enables both constant 
propagation and folding for UDFs with no additional effort. Since the 
entire UDF is now a single relational expression, SQL Server's existing scalar 
simplification mechanisms simplify the expression. \reffig{fig:slicing}(d) shows how the 
expression is simplified by evaluating both the predicate ($@x > 1000$) and then the 
string concatenation operation (`high' + ` value') at compile time, after 
propagating the constant `high'.

\subsection{Dead Code Elimination}
Lines of code that do not affect the result of a program are 
called \textit{dead code}. 
Dead code includes code that can never be executed 
(unreachable code), and code that only affects dead variables (assigned, 
but never read). 
As an example, suppose the following line of code was 
present in function \textit{total\_price} (\reffig{fig:udf-example}) between 
lines 3 and 4:

\centerline{\textbf{select }\textit{@t=count(*)} \textbf{from }\textit{orders} \textbf{where }\textit{o\_custkey=@key}}

The above line of code assigns the result of a query to a variable that is never
used, and hence it is dead code. In our experiments, we found many occurrences of 
dead code. As UDFs evolve and grow more complex, it becomes hard for 
developers to keep track of unused variables and code. 
%This leads to unnecessary execution of these instructions. 
Dead code can also be formed as a consequence of other optimizations. 
Dead code elimination is a 
technique to remove such code during compilation~\cite{Aho06}. Since UDFs are interpreted, 
most forms of dead code elimination are not possible.

Now let us consider how Froid handles this. Since the variable 
\textit{@t} is in the Write-Set of R1, it appears as an attribute of DT1. 
However, since it is never used, there will be no reference to \textit{DT1.t} 
in the final expression. Since there is an explicit projection 
on the \textit{returnVal} attribute, \textit{DT1.t} is like an 
attribute of a table that is not present in the final projection list of a 
query. Such attributes are aggressively removed by the optimizer using projection pushdown. 
Thereby, the entire sub-expression corresponding to the variable \textit{@t} gets 
pruned out, eliminating it from the final expression. 

\mysubsection{Summary}
We showed how Froid uses relational 
transformations to arrive at the same end result as that of applying 
compiler optimizations on imperative code. 
One might argue that compiler optimizations could be implemented for UDFs 
%directly instead of 
without using Froid's approach. However, that would only be a partial solution since it does 
not address inefficiencies due to iterative UDF invocation and serial plans. 

We conclude this section by highlighting two other aspects. First, 
the semantics of the \textit{Apply} operator allows the query optimizer to move and reuse
operations as necessary, while preserving correlation dependencies. 
This achieves the outcome of \textit{dependency-preserving statement reorderings} and 
\textit{common sub-expression elimination}~\cite{Aho06}, often used by optimizing compilers. 
Second, due to the way Froid is designed, these techniques are automatically
applied across nested function invocations, resulting in increased benefits due to 
\textit{interprocedural optimization}. 
%This is another reason that makes Froid widely 
%applicable and beneficial to real world UDFs. 

%\reffig{fig:slicing} also illustrates how the techniques described in this section work 
%together. The resulting program after slicing (\reffig{fig:slicing}(b)) is in a form
%suitable for applying constant propagation for the constant `high'. Then, constant
%folding is triggered, which performs the concatenation (\textit{`high' + `value'}) at 
%compile time (\reffig{fig:slicing}(c)). This results in 2 lines of dead code which 
%are eliminated, resulting in a constant return value (\reffig{fig:slicing}(d)). 

%\mysubsection{Interprocedural optimizations}
%We would like to highlight an important aspect of Froid. Due to the 
%way Froid is designed, all the above techniques are enabled across nested function 
%invocations. This is another reason that makes Froid widely applicable and 
%beneficial to real world UDFs. 

%Such interprocedural
%optimizations are generally quite expensive to implement in programming language 
%compilers, and hence many of these compiler optimizations are intraprocedural by default 
%\footnote{Most of the popular compilers do allow interprocedural optimizations; 
%however they need to be explicitly enabled, and are generally not enabled by default.}.

\section{Design and Implementation} \label{sec:impl}
In this section, we discuss key design choices, trade-offs,
and implementation details of the Froid framework.

\subsection{Cost-based Substitution} \label{subsec:cbi}
One of the first questions we faced while designing Froid was to decide whether
inlining of UDFs should be a cost-based decision. The answer to this question
influences the choice of whether substitution should be performed during Query
Optimization (QO) or during binding.

If inlining has to be a cost-based decision, it has to be performed during QO. 
If not, it can be done during binding. There are trade-offs to
both these design alternatives. One of the main advantages to doing this during binding
is that it is non-intrusive -- the QO and other phases of query processing require no
modifications. On the other hand, inlining during query optimization has the advantage
of considering the algebrized UDF as an alternative, and making a cost-based
decision of whether to substitute or not. 

In Froid, we chose to perform inlining during binding due to these reasons:
(a) Our experiments on real workloads showed that the inlined version performs better 
in almost all cases (see \refsec{sec:eval}), questioning the need for cost-based substitution.
(b) It is non-intrusive, requiring no changes to the query optimizer -- this is an 
important consideration for a commercial database system,
(c) Certain optimizations such as constant folding are performed during binding.
Inlining during QO would require re-triggering these mechanisms 
explicitly, which is not desirable.

\subsection{Imposing Constraints} \label{subsec:limits}
Although Froid improves performance in most cases, there are extreme 
cases where it might not be a good idea. Algebrization can  
increase the size and complexity of the resulting query (see \refsec{sec:applicability}). 
From our experiments, we found that transforming a UDF with thousands 
of lines of code may not always be desirable as it could lead to a query tree with tens of 
thousands of operators. Additionally, note that the query invoking the UDF might 
itself be complex as well (see \refsec{subsubsec:complexqudf}). Optimizing such a huge input tree makes the 
job of the query optimizer very hard. The space of alternatives to consider would 
increase significantly. 

To mitigate this problem, we have implemented a set of algebraic transformations that simplify 
the query tree reducing its size when possible. 
However, in some cases, the query tree may remain huge even after simplification. 
This has an impact on optimization time, and also on the 
quality of the plan chosen. Therefore, one of the constraints we imposed on 
Froid is to restrict the size of algebrized query tree. In turn, this restricts the 
size of UDFs that are algebrized by Froid. Based on our experiments, we found that 
except for a few extreme cases (see \refsec{subsubsec:factor}), imposing this 
constraint still resulted in significant performance gains.
%did not reduce the applicability of Froid, based on our experiments. 
%Note that 
%in our implementation, we have a set of algebraic transformations that simplify 
%the algebrized query tree thereby reducing its size whenever possible. Details are 
%omitted due to lack of space.

%, but it suffices to say that these transformations 
%are based on known algebraic simplifications. 

%TODO: minimizing algebrizer output - optimization. example in appendix?

\mysubsection{Nested and Recursive functions} \label{subsec:nestrec}
Froid's transformations can result in deep and complex trees 
(in the case of deeply nested function calls), or never terminate at all 
(in the case of recursive UDFs), if it is not managed appropriately. 
Froid overcomes this problem by controlling
%A common approach used by compilers, which we adopt, is the notion of `inlining depth'. 
%The inlining depth could be maintained and incremented every time a nested UDF is inlined. 
%With a threshold value for the maximum inlining depth, we can control the complexity
%of inlining deeply nested function calls. A more suitable approach to control 
%the complexity would be to control 
the inlining depth based on the size of the algebrized tree. This allows algebrization of
deeper nestings of smaller UDFs and shallow nestings of larger UDFs.
%We define a threshold for the maximum inlining depth. Therefore, Froid performs 
%inlinining up to a depth defined by the threshold, and stops. This approach is 
%common to both nested and recursive UDFs. 
%\mysubsection{Tuning and evaluation of nesting depth}
Note that if there is a deep nesting of large UDFs (or recursive UDFs), algebrizing 
a few levels might still leave UDFs in the query. This still is 
highly beneficial in terms of reducing 
function call overheads and enabling the choice of set-oriented plans, but 
it does not overcome the limitation on parallelism (\refsec{sec:udfperf}).

\subsection{Supporting additional languages} \label{sec:extend}
Relational databases allow UDFs and procedures to be written in imperative languages 
other than procedural SQL, such as C\#, Java, R and Python. Although the specific syntax 
varies across languages, they all provide constructs for common imperative operations such as 
variable declarations, assignments and conditional branching. Froid is 
an extensible framework, designed in a way that makes it straightforward to incrementally add 
support for more languages and imperative constructs. 

Froid models each imperative construct as a class that encapsulates the logic for algebrization 
of that construct. Therefore, adding support for additional languages only 
requires (a) plugging in a parser for that language and (b) providing a 
language-specific implementation for each supported construct. The framework itself is 
agnostic to the language, and hence remains unchanged. 
As long as the UDF is 
written using supported constructs, Froid will be able to algebrize them as 
described in this paper. 

Note that while translating from a different 
language into SQL, data type semantics need to be taken into account to ensure 
correctness. Data type semantics vary across languages, and translating
to SQL might lead to loss of precision, and sometimes different results.

\subsection{Implementation Details}
We now briefly discuss some special cases and other implementation details.

\noindent\textbf{Security and Permissions}
Consider a user that does not have \textit{execute} permissions on the UDF, but has 
\textit{select} permissions on the referenced tables. Such a user will be able to run 
an inlined query (since it no longer references the UDF), even though it should 
be disallowed. To mitigate this issue, Froid enlists the UDF for permission 
checks, even if it was inlined. Conversely, a user may have \textit{execute} 
permission on the UDF, but no \textit{select} permissions on the referenced tables. 
In this case, by inlining, that user is unable to run the query even though it should 
be allowed. Froid handles this similar to the way view permissions are handled.
% currently in SQL Server.

\mysubsection{Plan cache implications}
Consider a case where a user with administrative privileges runs a query 
involving this UDF, and consequently the inlined plan is now cached. 
Subsequently, if a user without UDF \textit{execute} permissions but 
with \textit{select} permissions on the underlying tables runs the same query, the 
cached plan will run successfully, even though it should not. 
Another implication is related to managing metadata version changes and cache 
invalidation. Consider the case as described above, where an inlined plan is 
cached. Now, if the user alters or drops the UDF, the UDF is changed or no longer 
available. Therefore, any query that referred to this UDF should be removed from 
the plan cache. Both these issues are solved by enlisting the UDF in 
schema and permission checks, even if it was algebrized.   

\mysubsection{Type casting and conversions}
SQL Server performs implicit type conversions and casts in many cases when the datatypes of 
parameters and return expressions are different from the declared types. In 
order to preserve the semantics as before, \textsl{Froid} explicitly inserts appropriate type casts 
for actual parameters and the return value. 

\mysubsection{Non-deterministic intrinsics}
UDFs may invoke certain non-deterministic functions such as GETDATE().
Inlining such UDFs might violate the user's intent since it may invoke
the intrinsic function once-per-query instead of once-per-tuple. Therefore, we 
disable transforming such UDFs.

%TODO: design alternatives - create time algebrization, 
%TODO: pure functions
\section{Evaluation} \label{sec:eval}
\begin{figure*}[t!]
	\minipage{0.65\columnwidth}
	\includegraphics[width=\linewidth]{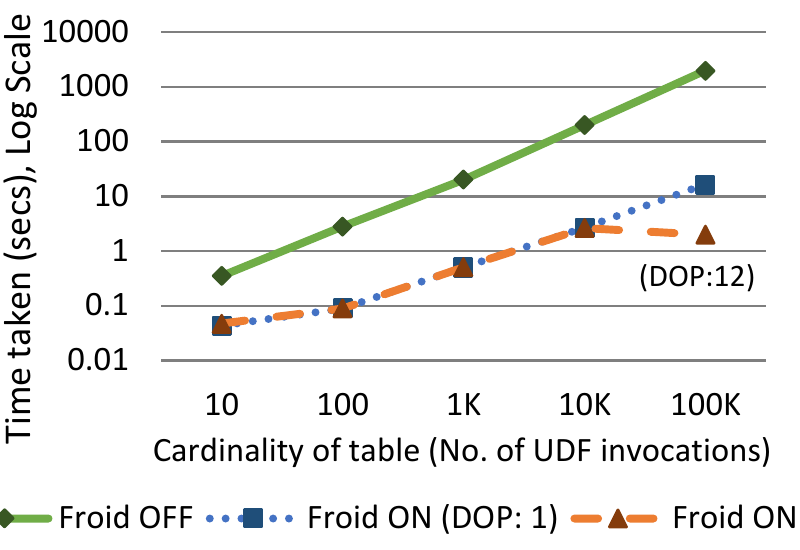}
	\caption{Varying the number of UDF invocations}\label{fig:varycard}
	\endminipage
	\hspace{2mm}
	\minipage{0.7\columnwidth}
	\includegraphics[width=\linewidth]{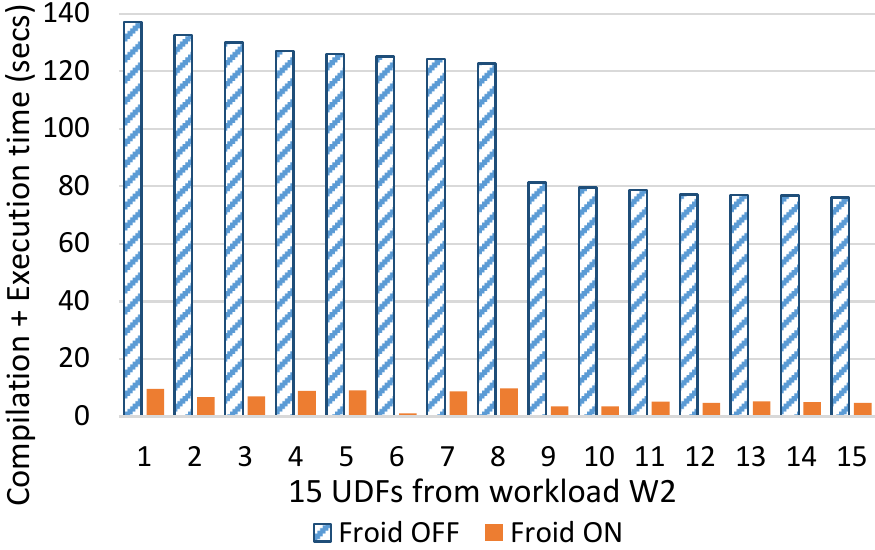}
	\caption{Elapsed time for Compilation and execution (using cold plan cache)}\label{fig:comptime}
	\endminipage
	\hspace{2mm}
	\minipage{0.7\columnwidth}
	\includegraphics[width=\linewidth]{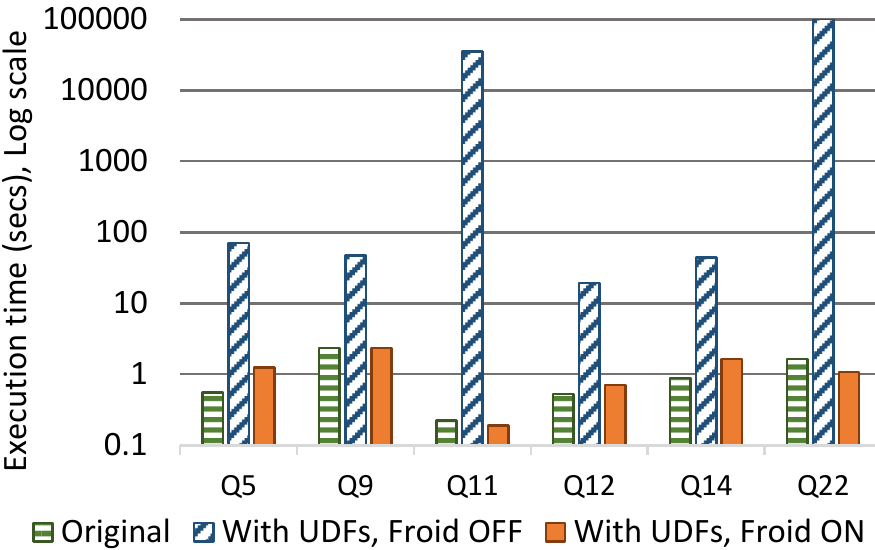}
	\caption{TPC-H queries using UDFs}\label{fig:tpch-udf}
%	\caption{CPU Time comparison}\label{fig:cputime}
	\endminipage
\end{figure*}
We now present some results of our 
evaluation of Froid on several workloads and configurations.
%disk-based row stores, column stores and memory-optimized 
%configurations. We consider warm and cold cache scenarios, 
%and present the benefits due to Froid on both performance and resource consumption. 
Froid is implemented in SQL Server 2017 in about 1.5k 
lines of code. %, and supports the following constructs.
%\begin{itemize}[leftmargin=*,noitemsep]
%	\item \textbf{DECLARE, SET:} Variable declaration and assignments.
%	\item \textbf{SELECT:} SQL query, with multiple variable assignments.
%	\item \textbf{IF/ELSE:} Branching with arbitrary levels of nesting.
%	\item \textbf{UDF:} Nested/recursive function calls.
%	\item \textbf{RETURN:} Single or multiple return statements.
%	\item \textbf{Others:} Relational operations such as EXISTS, ISNULL.	
%\end{itemize}
For our experiments, SQL Server 2017 with Froid was run on 
Windows Server 2012(R2). The machine was equipped with Intel Xeon X5650 2.66 
Ghz CPU (2 processors, 6 cores each), 96 GB of RAM and SSD-backed storage.

\subsection{Applicability of Froid} \label{sec:applicability}

\begin{table}
	\small
	\centering	
	\begin{tabular}{ |c|c|c| } 
		\hline
		\textbf{Workload} &  \textbf{W1} & \textbf{W2} \\
		\hline 
		\text{Total \# of scalar UDFs} & 178 & 93 \\
		\hline
		\text{\# UDFs optimizeable by Froid} & 151 (85\%)  & 86 (92.5\%) \\
		\hline
		\text{UDF lines of code (avg,min,max)} & (21,6,113)  & (26,7,169) \\
		\hline
	\end{tabular}
	\caption{Applicability of Froid on two customer workloads}	
	\label{tab:applicability} 
	\vspace{-1.5em}
\end{table}

We have analyzed several customer workloads from Azure SQL Database to measure the 
applicability of Froid with its currently supported constructs. We are primarily
interested in databases that make good use of UDFs and hence, we considered 
the top 100 databases in decreasing order of the number of UDFs present in 
them. Cumulatively, these 100 databases had 85329 scalar UDFs, out of
which Froid was able to handle 51047 (59.8\%). The UDFs that could not be 
transformed contained constructs not supported by Froid. 
We also found that there are 10526 customer 
databases with more than 50 UDFs each, where Froid can inline more than 70\% of 
the UDFs. The sizes of these UDFs range from a single line to 1000s
of lines of code. These numbers clearly demonstrate the wide applicability of Froid.

In order to give an idea of the kinds of UDFs that are in these propreitary workloads, we have 
included a set of UDFs in \refsec{sec:udfexamples}. % of our technical report~\cite{FroidTechRep}. 
These UDFs have been modified to preserve anonymity, while retaining program structure. 
We have randomly chosen two customer workloads (referred to as W1 and W2) for 
deeper study and performance analysis. The UDFs have been used with no modifications, 
and there were no workload-specific techniques added to Froid. As summarized in Table~\ref{tab:applicability}, 
Froid is able to transform a large fraction of UDFs in these workloads (85\% and 92.5\%). 
As described in~\refsec{sec:impl}, UDF algebrization results in larger query trees as 
input to query optimization. 
%Out of the 86 inline-able UDFs in workload W2, 30 had a nesting depth less than 3, and 56 UDFs had depths greater than 3. 
The largest case in W2 resulted in more than 300 imperative statements being transformed into a 
single expression, having more than 7000 nodes. 
Note that this is prior to optimizations described in~\refsec{sec:compileropt}. 
This illustrates the complexity of UDFs handled by Froid.

%\reffig{fig:udfsizes} shows this result for workload W2. The x-axis shows the
%number of statements in the UDF including nested UDFs, and the y-axis shows
%the size of the inlined query tree (in terms of the number of nodes).
%Each column in \reffig{fig:udfsizes} represents a UDF, and the colors indicate
%whether the nesting depth of the UDF was $<=3$ (blue) or $>3$ (orange).
%
%We observe that the size of the inlined tree can go up to thousands of nodes,
%even when max inlining depth is set to 3. For all cases where the nesting
%depth was $>3$, we stop inlining at level 3 in our experiments. \reffig{fig:udfsizes} 
%also gives an idea of the complexity of UDFs inlinable by Froid.

\subsection{Performance improvements} \label{subsec:perf}

We now present a performance evaluation of Froid on workloads W1 and W2. Since our primary 
focus is to measure the performance of UDF evaluation, the queries that invoke 
UDFs are kept simple so that UDF execution forms their main component. Evaluation of
complex queries with UDFs is considered in \refsec{subsubsec:complexqudf}.

\subsubsection{Number of UDF invocations}

%\onecolfigure
%{figs/varycard}
%{Varying the number of UDF invocations}
%{fig:varycard}

The number of times a UDF is invoked as part of a query has a significant impact
on the overal query performance. In order to compare the relationship between 
the number of UDF invocations and the corresponding performance gains, we consider
a function F1 (which in turn calls another function F2).
%Appendix~\ref{app:udf-examples}. 
F1 and F2 are functions adapted from workload W1, and their definitions are given in \refsec{sec:udfexamples}. 
We use a simple query to invoke this UDF, of the form\\
\centerline{\textbf{select }\textit{dbo.F1(T.a, T.b)} \textbf{from }\textit{T}}\\
Since the UDF is invoked for every tuple in \textit{T}, we can control the number
of UDF invocations by varying the cardinality of \textit{T}. \reffig{fig:varycard}
shows the results of this experiment conducted with a warm cache. The x-axis 
denotes the cardinality of table \textit{T} (and hence the number of UDF invocations),
and the y-axis shows the time taken in seconds, in log scale. Note that in this 
experiment, the time shown in the y-axis does not include query compilation time, 
since the query plans were already present in the cache.

We vary the cardinality of \textit{T} from 10 to 100000. With Froid disabled,
we observe that the time taken grows with cardinality (the solid line
in \reffig{fig:varycard}). With Froid enabled, we see an improvement of one to 
three orders of magnitude (the dashed line). The advantages start to be noticeable 
right from a cardinality of 10.

\onecolfigure
{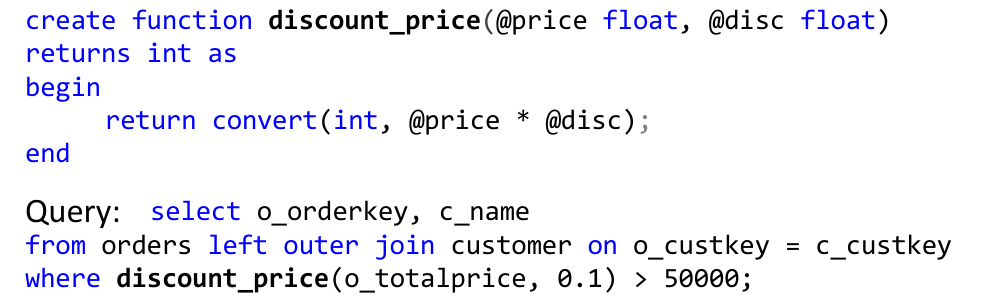}
{Example for \refsec{subsec:cci}}
{fig:cci}
%\twocolfigure
%{figs/tmw-factor}
%{Improvement due to Froid for UDFs in workload W1}
%{fig:w1factor}
%\onecolfigure
%{figs/crm-factor-1col}
%{Improvement due to Froid on workload W2}
%{fig:w2factor}

\begin{figure*}[t!]
	\minipage{1.35\columnwidth}
	\includegraphics[width=\linewidth]{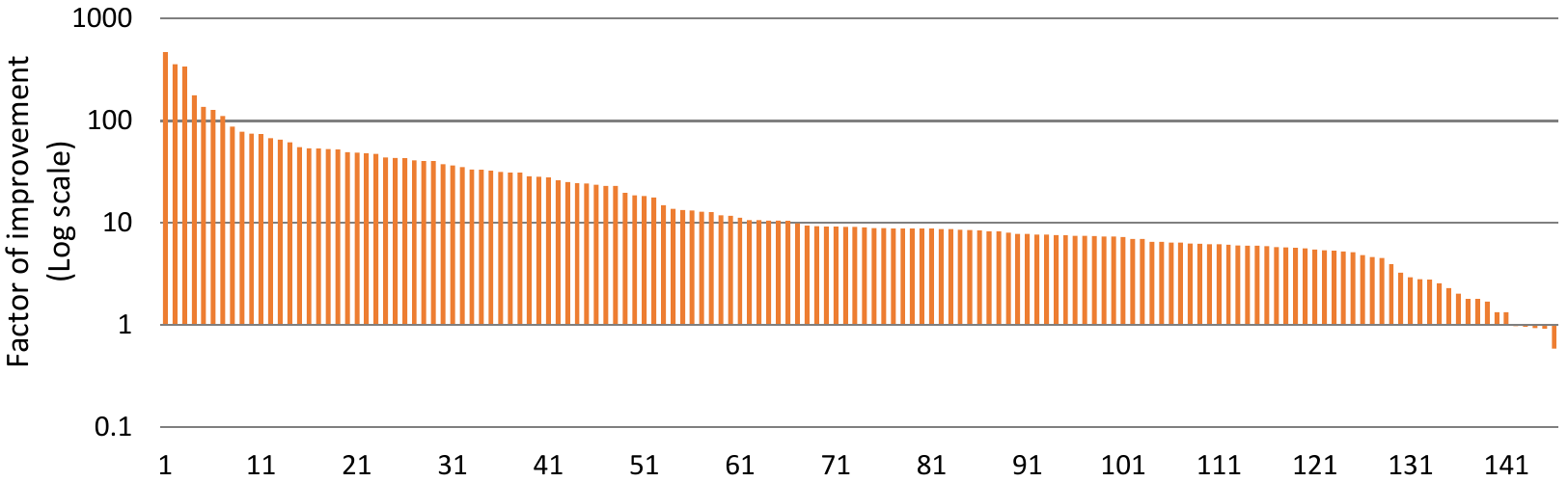}
	\caption{Improvement for UDFs in workload W1}\label{fig:w1factor}
	\endminipage
	\hspace{2mm}
	\minipage{0.7\columnwidth}
	\includegraphics[width=\linewidth]{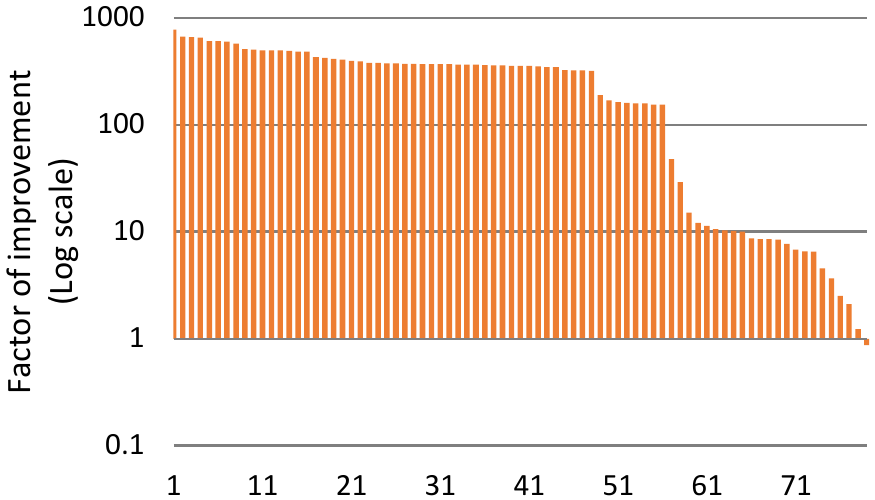}
	\caption{Improvement for UDFs in workload W2}\label{fig:w2factor}
	\endminipage
\end{figure*}

\subsubsection{Impact of parallelism}
As described in this paper, Froid brings the benefits of set-oriented plans,
compiler optimizations, and parallelism to UDFs.
In order to isolate the impact of parallelism from
the rest of the optimizations (since enabling parallelism is a by-product of 
Froid's transformations), we conducted experiments where we enabled Froid 
but limited the Degree Of Parallelism (DOP).
The dotted line in \reffig{fig:varycard} shows a result of this experiment.
It includes all the optimizations of Froid, but forces the DOP to 1 using a query 
hint. For this particular UDF, SQL Server switches to a 
parallel plan when the cardinality of the table is greater than 10000 (indicated 
by the dashed line). The key observation we make here is that even without parallelism, 
Froid achieves improvements up to two orders of magnitude.

\subsubsection{Compile time overhead} \label{subsubsec:comptime}
Since Froid is invoked during query compilation, there could be an increase in 
compilation time. This increase is not a concern as it is offset by the 
performance gains achieved. To quantify this, we measured the total elapsed time
including compilation and execution by clearing the plan cache before running
queries. This keeps the buffer pool warm, but the plan cache cold.
The results of this experiment on 15 randomly chosen UDFs (sorted in descending 
order of elapsed time) of workload W2 are shown in \reffig{fig:comptime}. The 
y-axis shows total elapsed time which includes compilation and execution.
We observe gains of more than an order of magnitude for all these UDFs. 
Note that the compilation time of each of these UDFs is less than 10 seconds.

%Also, due to caching of compiled
%plans, this overhead is negligible in most workloads.
%A possible improvement here is to perform the
%transformations of Froid at function creation time, and store the 
%relational equivalent of the UDF along with its metadata. With that, query 
%compilation overheads can be eliminated. 

\subsubsection{Complex Analytical Queries With UDFs} \label{subsubsec:complexqudf}
In the above experiments, we kept the queries simple so that the UDF forms the 
main component. To evaluate Froid in situations where the queries invoking UDFs 
are complex, we considered TPC-H~\cite{TPCH} queries, and looked for 
opportunities where parts of queries could be expressed using scalar UDFs. 
We extracted several UDFs and then modified the queries to use these UDFs. 
The UDF definitions and rewritten queries are given in \refsec{sec:tpch-udfs-list}. %our technical report~\cite{FroidTechRep}. 
\reffig{fig:tpch-udf} shows the results on a 10GB TPC-H dataset with warm cache for 
6 randomly chosen queries. For each query, we show the time taken for (a) the 
original query (without UDFs), (b) the rewritten query with UDFs (with Froid OFF), 
and (c) the rewritten query with Froid ON. 

Observe that for all queries, Froid leads to improvements of multiple orders of 
magnitude (compare (b) vs. (c)). We also see that in most cases, there is 
no overhead to using UDFs when Froid is enabled (see (a) vs. (c)). These 
improvements are the outcome of all the optimizations that are enabled by Froid.
For some queries (eg. Q5, Q14), there is a small overhead when compared with 
original queries. There are also cases (eg. Q11, Q22) where Froid does slightly better 
than the original. An analysis of query plans revealed that these are due to 
slight variations in the chosen plan as a result of Froid's transformations. 
The details of plan analysis are beyond the scope of this paper.
%This is in cases where the optimizer is unable to 
%push down some predicates due to the expressions generated by Froid. 

\subsubsection{Factor of improvement} \label{subsubsec:factor}
We now consider the overall performance gains achieved due to Froid on workloads W1 
and W2 (row store), shown in Figures \ref{fig:w1factor} and \ref{fig:w2factor}. 
The size of table \textit{T} was fixed at 100,000 rows, and queries were run with 
warm cache (averaged over 3 runs). In these figures, 
UDFs are plotted along the x-axis, ordered by the observed improvement with Froid 
(in descending order). The y-axis shows the factor of improvement (in log scale). 
We observe improvements in the range of 5x-1000x across both workloads. In total, 
there were 5 UDFs that showed no improvement or performed slightly worse due to 
Froid. One of the main reasons for this was the presence of complex recursive 
functions. These can be handled by appropriately tuning the constraints as 
described in \refsec{subsec:limits}. UDFs that invoke expensive TVFs was another 
reason. Since our implementation currently does not handle TVFs, such UDFs do 
not benefit from Froid.

\subsubsection{Columnstore indexes} \label{subsec:cci}

\begin{table}
	\small
	\centering		
	%	\vspace{5mm}
	\begin{tabular}{ |c|c|c| } 
		\hline
		\textbf{Configuration} & \textbf{Froid OFF} & \textbf{Froid ON} \\
		\hline
		\textbf{Row store} & 24241 ms & 822 ms \\
		\hline
		\textbf{Column store} & 19153 ms & 155 ms\\
		\hline
	\end{tabular}
	\caption{Benefits of Froid on row and column stores (total elapsed time with cold cache) for the example in \reffig{fig:cci} .}	
	\label{tab:cci} 
	\vspace{-1em}
\end{table}

%We now consider the impact of using Froid in conjunction with columnstores, 
%a technology for managing data using a columnar data format. 
We now present the results of our experiments on column stores.
Column-stores
achieve better performance because of high compression rates, smaller memory footprint, and
batch execution~\cite{CSIG}. 
%Batch mode execution refers to the mechanism of 
%processing a set of rows together. This amortizes the overhead costs over all the 
%rows in a batch, thereby significantly reducing costs. Batch mode operates on 
%compressed data when possible, speeding up execution of analytics queries significantly.
%\onecolfigure
%{figs/cci}
%{Example for \refsec{subsec:cci}}
%{fig:cci}
%Not all query execution operators can be executed in batch mode. 
However, encapsulating aggregations and certain other operations inside a UDF prevents
the optimizer from using batch mode for those operations. Froid brings
the benefits of batch mode execution to UDFs. Consider a
simple example based on the TPC-H schema as shown in \reffig{fig:cci}.
The results of running this on a TPC-H 1GB database with a cold cache are shown in 
Table~\ref{tab:cci}.

For this example, without Froid, using a clustered columnstore index (CCI) 
led to about 20\% improvement 
in performance over row store. With Froid, however, we get about
5x improvement in performance by using column store over row store. Along with
other reasons, the fact that the predicate and discount computation can now
happen in batch mode contributes to the performance gains.

\subsubsection{Natively compiled queries and UDFs}
\begin{table}
	\small
	\centering	
	\begin{tabular}{ |c|c|c| } 
		\hline
		\textbf{Configuration} & \textbf{Froid OFF} & \textbf{Froid ON} \\
		\hline		
		Query and UDF interpreted & 41729 ms & 2056 ms\\
		\hline
		Interpreted query, native UDF & 27376 ms & NA \\
		\hline
		Native query, native UDF & 9230 ms & 2005 ms \\
		\hline
	\end{tabular}
	\caption{Benefits of Froid with native compilation (total elapsed time with warm cache) for the UDF in~\protect\cite{RBAR}.}	
	\label{tab:hk} 
	\vspace{-1em}
\end{table}

Hekaton, the memory-optimized OLTP engine in SQL Server performs native 
compilation of procedures~\cite{HekaSig}, which allows 
more efficient query execution than interpreted T-SQL~\cite{NCSP}. Due to its 
non-intrusive design, Froid seamlessly integrates with Hekaton and provides 
additional benefits. For this expriment, 
we considered the UDFs (\textit{dbo.FarePerMile}) used in an MSDN article 
about native compilation~\cite{RBAR} (the UDFs are reproduced in \refsec{sec:nativeexamples}). 
We considered a memory optimized table with 3.5 million rows and 25 columns, 
with a CCI. The results of this experiment %for 3 possible configurations are 
are shown in Table~\ref{tab:hk}.

First, in the classic mode of interpreted T-SQL, we see a 20x improvement 
due to Froid. Next, we natively compiled the UDF, but ran the query in interpreted 
mode. This results 
in a 1.5x improvement compared to the fully interpreted mode with Froid disabled. 
Froid is not applicable here since a compiled 
module cannot be algebrized. 

Finally, we natively compiled both the UDF and the 
query, and ran it with and without Froid enabled. With Froid disabled, we see 
the full benefits of native compilation over interpreted mode, with a 4.5x 
improvement. With Froid enabled, we get the combined benefits of algebrization 
and native compilation. Froid first inlines the UDF, and then
the resulting query is natively compiled, \textit{giving an additional 
4.6x improvement over native compilation}.
Although native compilation makes UDFs faster, the benefits are limited as 
the query still invokes the UDF for each tuple. Froid removes this fundamental 
limitation and hence combining Froid with native compilation leads to more gains.

\subsection{Resource consumption}
\onecolfigure
{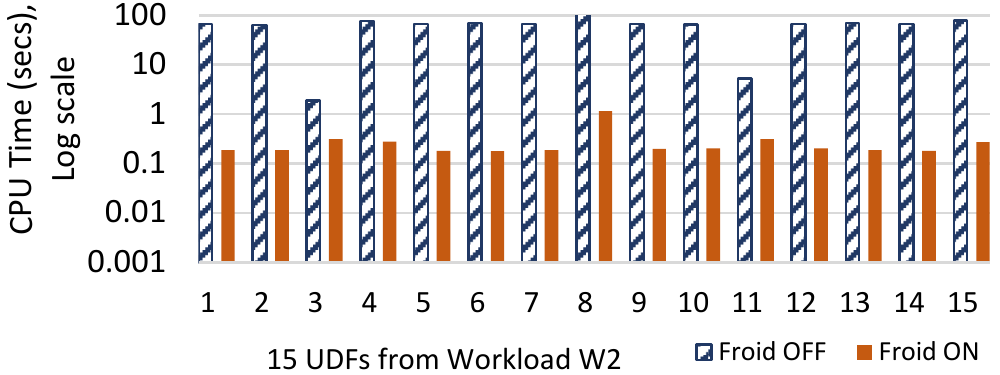}
{CPU time comparison}
{fig:cputime}
\onecolfigure
{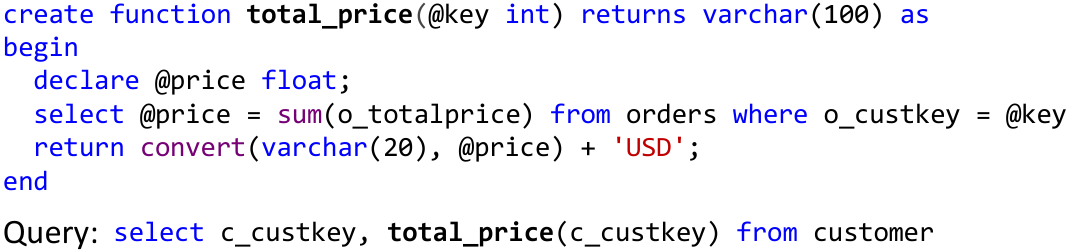}
{Example for I/O measurements}
{fig:io}

In addition to significant performance gains, our techniques offer an additional 
advantage -- they significantly reduce the resources consumed by such queries. 
The reduction in CPU time due to Froid is shown in \reffig{fig:cputime}. Due to 
lack of space, we show the results for a randomly chosen subset of UDFs from 
workload W2; the results were similar across all the workloads we evaluated. Observe 
that Froid reduces the CPU time by 1-3 orders of magnitude for all UDFs. This 
reduction is due to elimination of expensive context-switches (see 
\refsec{subsec:udfeval}), and also due to optimizations such as set-oriented 
evaluation, folding and slicing. % (see \refsec{sec:compileropt}).

Due to the above-mentioned reasons, Froid also reduces I/O costs. The I/O metric 
is dependent upon the nature of operations in the UDF. For UDFs that perform 
data access, our transformations will lead to reductions in logical reads as it 
avoids repetition of data access for every invocation of the UDF. Consider a 
simple UDF such as the one in \reffig{fig:io}. With Froid, the query requires 
about 3300 logical reads, whereas without Froid, it issued close to 5 million
logical reads on a 1GB TPC-H dataset with cold cache. Such improvements lead to 
significant cost savings for our customers, especially for users of cloud 
databases, since they are billed for resources they consume.

\section{Real-World UDF Examples} \label{sec:udfexamples}
In this section, we provide some examples adapted from real world UDFs. 
They give an idea of the kinds of UDFs that 
are commonly encountered in practice. These have been modified 
to preserve anonymity, while retaining program structure.
All the UDFs given in this section are inlineable by Froid. As it
can be seen, Froid can handle a fairly large class of UDFs encountered
in practice. 

%Due to lack of space, we omit listing larger functions which are quite 
%commonly encountered.
%\mysubsection{1. Single statement functions}\\
%These are commonly used functions for mathematical calculations,
%date and string manipulations.\\
%\noindent UDF1: The below function returns the day of week for a given \textit{datetime} object.
{\scriptsize
\begin{Verbatim}[obeytabs, tabsize=2]
create function dbo.F1(@p1 int, @p2 int) 
returns bit as 
begin
	if EXISTS 
		(SELECT 1 FROM View1 WHERE col1 = 0
		AND col2 = @p1
		AND ((col2 = 2) OR (col3 = 2))
		AND dbo.F2(col4,@p2,0)=1 AND dbo.F2(col5,@p2,0)=1
		AND dbo.F2(col6,@p2,0)=1 AND dbo.F2(col7,@p2,0)=1
		AND dbo.F2(col8,@p2,0)=1 AND dbo.F2(col9,@p2,0)=1
		AND dbo.F2(col10,@p2,0)=1 AND dbo.F2(col11,@p2,0)=1
		AND dbo.F2(col12,@p2,0)=1 AND dbo.F2(col13,@p2,0)=1
		AND dbo.F2(col14,@p2,0)=1 AND dbo.F2(col15,@p2,0)=1)
	  return 1
	return 0
end
\end{Verbatim}
}
\divider
{\scriptsize
\begin{Verbatim}[obeytabs, tabsize=2]
create function dbo.F2(@p1 int,@p2 int, @flag1 int = 0) 
returns bit AS 
begin
	DECLARE @Flag bit	
	IF @flag1=0	BEGIN
		IF EXISTS (SELECT 1 FROM Table1 
				WHERE col1=@p1 AND col2=@p2) 
		OR @p1 Is Null
			SET @Flag= 1
		ELSE  SET @Flag= 0
	END    
	ELSE BEGIN
		IF EXISTS (SELECT 1 FROM Table1 T1 
				INNER JOIN Table2 T2 
				ON T1.col1=T2.col2 
				WHERE T2.col2=@p1 AND T1.col2=@p2) 
		OR @p1 Is Null
			SET @Flag= 1
		ELSE SET @Flag= 0
	END
	return @Flag
end
\end{Verbatim}
}
\divider
{\scriptsize
\begin{Verbatim}[obeytabs, tabsize=2]
create function dbo.DayOfWeek(@d datetime) returns int as
begin
	return (DATEPART(dw, @d) + @@DATEFIRST -1) % 7
end	
\end{Verbatim}	
}
\divider
%\mysubsection{2. Functions with straight-line code}\\
%\noindent UDF2: The below function returns the beginning of the hour in UTC
%for a given \textit{datetime} object.
{\scriptsize
\begin{Verbatim}[obeytabs, tabsize=2]
create function dbo.BeginOfHour(@d datetime)
returns datetime as
begin
	declare @DayBeginUTC datetime
	set @DayBeginUTC = convert(datetime, convert(nvarchar, @d, 112))
	return dateadd(hh, datepart(hh, @d), @DayBeginUTC)
end
\end{Verbatim}	
}
\divider
%\vspace{2mm}
%\noindent\rule{90mm}{.1pt}
%\noindent UDF3: The below function returns the beginning of the month in UTC
%for a given \textit{datetime} object.
{\scriptsize
\begin{Verbatim}[obeytabs, tabsize=2]
create function BeginOfMonth(@d datetime) returns datetime as
begin
	declare @DayUserLocal datetime, @DayFirst datetime
set @DayUserLocal = dbo.UTCToLocalTime(@d)
	declare @m = datepart(mm, @DayUserLocal)
	set @DayFirst = dbo.FirstDayOfMonth(@DayUserLocal, @m)  
	return dbo.LocalTimeToUTC(@DayFirst)
end
\end{Verbatim}	
}
\divider
%\mysubsection{4. Functions with conditional branching and multiple return statements}\\
%\noindent UDF4: The below function formats the range for a report.
{\scriptsize
\begin{Verbatim}[obeytabs, tabsize=2]
CREATE  FUNCTION dbo.RptBracket(@MyDiff int, @NDays int)
RETURNS nvarchar(10) AS
BEGIN
	if(@MyDiff >= 5*@NDays)
	begin
		RETURN ( Cast(5 * @NDays as nvarchar(5)) + N'+')
	end

	RETURN ( Cast(Floor(@MyDiff / @NDays) * @NDays as nvarchar(5)) 
		+ N' - ' 
		+ Cast(Floor(@MyDiff / @NDays + 1) * @NDays - 1 as nvarchar(5)))
END
\end{Verbatim}	
}
\divider
%\vspace{2mm}
%\noindent UDF5: This function returns the first day of the month for the given day.
{\scriptsize
\begin{Verbatim}[obeytabs, tabsize=2]
create function dbo.FirstDayOfMonth (@d datetime, @Month int)
returns datetime as
begin
	declare @Result datetime
	set @Result = dateadd( day, 1 - datepart( day, @d ), @d )
	if datepart( month, @Result ) <> datepart( month, @d ) 
		set @Result = NULL
		
	declare @mdiff int = @Month - datepart(mm, @Result);
	set @Result = dateadd( mm, @mdiff, @Result)
	return (convert(datetime, convert(nvarchar, @Result, 112)))
end
\end{Verbatim}	
}
\divider
%\noindent UDF6: This function accepts a version in \textit{varchar} and converts it into float.
{\scriptsize
\begin{Verbatim}[obeytabs, tabsize=2]
create function dbo.VersionAsFloat(@v nvarchar(96))
returns float as
begin
	if @v is null return null
	declare @first int, @second int;
	declare @major nvarchar(6), @minor nvarchar(10); 
	
	set @first = charindex('.', @v, 0);
	if @first = 0 
		return CONVERT(float, @v);

	set @major = SUBSTRING(@v, 0, @first);	
	set @second = charindex('.', @v, @first + 1);
	if @second = 0
		set @minor=SUBSTRING(@v, @first+1, len(@v)-@first)
	else
		set @minor=SUBSTRING(@v, @first+1, @second-@first-1);
	
	set @minor = CAST(CAST(@minor AS int) AS varchar);	
	return CONVERT(float, @major + '.' + @minor);
end
\end{Verbatim}
}
\divider
%\noindent UDF7: This function tests whether the query is being run by a user
%with sufficient privileges and returns the corresponding businessGUID.
{\scriptsize
\begin{Verbatim}[obeytabs, tabsize=2]
create function dbo.fn_FindBusinessGuid()
returns uniqueidentifier as
begin
	declare @userGuid uniqueidentifier
	declare @businessguid uniqueidentifier
	
	if (is_member('SomeRole') | is_member('SomeGroup')) = 1
	begin
		select @userGuid = cast(context_info() as uniqueidentifier)		
		if @userGuid is not null
		begin		
			select @businessguid = s.col4
			from T1 s
			where s.col1 = @userGuid
			return @businessguid
		end
	end
	
	select @businessguid = s.col3
	from T1 s
	where s.col1 = SUSER_SNAME()
	return @businessguid
end
\end{Verbatim}
}
\divider
%\noindent UDF8: This function tests whether the query is being run by a user
%with sufficient privileges and returns the corresponding userGUID.
{\scriptsize
\begin{Verbatim}[obeytabs, tabsize=2]
create function dbo.fn_FindUserGuid()
returns uniqueidentifier as
begin
	declare @userGuid uniqueidentifier	
	if (is_member('AppReaderRole') | is_member('db_owner')) = 1
	begin
		select @userGuid = cast(context_info() as uniqueidentifier)
	end
		
	if @userGuid is null
	begin
		select @userGuid = s.SystemUserId
		from SystemUserBase s
		where s.DomainName = SUSER_SNAME()	
	end
	return @userGuid
end
\end{Verbatim}
}
\divider
\section{Natively compiled UDFs} \label{sec:nativeexamples}
These UDFs are borrrowed from an MSDN article~\cite{RBAR} about the benefits 
of Hekaton. As mentioned earlier, Hekaton, the memory-optimized OLTP engine 
in SQL Server performs native compilation of procedures~\cite{HekaSig}, which 
allows faster data access and more efficient query execution than interpreted 
T-SQL~\cite{NCSP}. We have used the following UDFs along with Froid, to 
measure the additional benefits that we achieve using our techniques.

Here is the simple UDF in T-SQL interpreted form:
{\scriptsize
	\begin{Verbatim}[obeytabs, tabsize=2]
	CREATE FUNCTION dbo.FarePerMile ( @Fare MONEY, @Miles INT )
	RETURNS MONEY
	WITH SCHEMABINDING
	AS
	BEGIN
		DECLARE @retVal MONEY = ( @Fare / @Miles );	
		RETURN @retVal;
	END;
	
	\end{Verbatim}
}
\noindent Here is the simple UDF written as a native compiled version:

{\scriptsize
	\begin{Verbatim}[obeytabs, tabsize=2]
	CREATE FUNCTION dbo.FarePerMile_native (@Fare money, @Miles int)
	RETURNS MONEY
	WITH NATIVE_COMPILATION, SCHEMABINDING, EXECUTE AS OWNER
	AS
	BEGIN ATOMIC
	WITH (TRANSACTION ISOLATION LEVEL = SNAPSHOT, LANGUAGE = N’us_english’)
	
		DECLARE @retVal money = ( @Fare / @Miles)
		RETURN @retVal
	END	
	\end{Verbatim}
}

\section{TPC-H Queries with UDFs} \label{tpch-udfs-list}
In this section, we first
show some scalar UDFs extracted from TPC-H queries, and then show the queries rewritten to use
these scalar UDFs. Observe that there are some UDFs that are used in multiple queries,
highlighting the benefits of code reuse due to the use of UDFs. Without Froid,
these rewritten queries with UDFs exhibit poor performance as shown in \refsec{sec:eval}.
 
\subsection{Scalar UDF Definitions}
{\scriptsize
\begin{Verbatim}[obeytabs, tabsize=2]
create function dbo.discount_price(@extprice decimal(12,2), 
																	@disc decimal(12,2)) 
returns decimal(12,2) as
begin
	return @extprice*(1-@disc);
end
\end{Verbatim}
}
\divider

{\scriptsize
\begin{Verbatim}[obeytabs, tabsize=2]
create function dbo.discount_taxprice(@extprice decimal(12,2), 
																		@disc decimal(12,2), 
																		@tax decimal(12,2)) 
returns decimal(12,2) as
begin
	return dbo.discount_price(@extprice, @disc) * (1+@tax);	
end
\end{Verbatim}
}
\divider

{\scriptsize
\begin{Verbatim}[obeytabs, tabsize=2]
create function dbo.profit_amount(@extprice decimal(12,2), 
										@discount decimal(12,2), 
										@suppcost decimal(12,2), 
										@qty int) 
returns decimal(12,2) as
begin
	return @extprice*(1-@discount)-@suppcost*@qty;
end
\end{Verbatim}
}
\divider

{\scriptsize
\begin{Verbatim}[obeytabs, tabsize=2]
create function dbo.isShippedBefore(@shipdate date, 
																	@duration int, 
																	@stdatechar varchar(10)) 
returns int as
begin
	declare @stdate date = cast(@stdatechar as date);
	declare @newdate date =  dateadd(dd, @duration, @stdate);
	if(@shipdate > @newdate)
		return 0;
	return 1;
end
\end{Verbatim}
}
\divider

{\scriptsize
\begin{Verbatim}[obeytabs, tabsize=2]
create function dbo.checkDate(@d varchar(10), 
																@odate date, 
																@shipdate date) 
returns int as
begin
	if(@odate < @d AND @shipdate > @d)
		return 1;
	return 0;
end
\end{Verbatim}
}
\divider

{\scriptsize
\begin{Verbatim}[obeytabs, tabsize=2]
create function dbo.q3conditions(@cmkt varchar(10), 
																@odate date, 
																@shipdate date) 
returns int as
begin
	declare @thedate varchar(10) = '1995-03-15';
	if(@cmkt <> 'BUILDING') 
		return 0;
	if(dbo.checkDate(@thedate, @odate, @shipdate) = 0)
		return 0;
	if(dbo.isShippedBefore(@shipdate, 122, @thedate) = 0)
		return 0;
	return 1;
end
\end{Verbatim}
}
\divider

{\scriptsize
\begin{Verbatim}[obeytabs, tabsize=2]
create function dbo.q5Conditions(@rname char(25), 
																@odate date) 
returns int as 
begin
	declare @beginDatechar varchar(10) = '1994-01-01';
	declare @beginDate date = cast(@beginDatechar as date);
	declare @newdate date;

	if(@rname <> 'ASIA') 
		return 0;
	if(@odate < @beginDate)
		return 0;

	set @newdate = DATEADD(YY, 1, @beginDate);
	if(@odate >= @newdate)
		return 0;

	return 1;
end
\end{Verbatim}
}
\divider

{\scriptsize
\begin{Verbatim}[obeytabs, tabsize=2]
create function dbo.q6conditions(@shipdate date, 
																@discount decimal(12,2), 
																@qty int) 
returns int as
begin
	declare @stdateChar varchar(10) = '1994-01-01';
	declare @stdate date = cast(@stdateChar as date);
	declare @newdate date = dateadd(yy, 1, @stdate);

	if(@shipdate < @stdateChar)
		return 0;

	if(@shipdate >= @newdate)
		return 0;

	if(@qty >= 24)
		return 0;

	declare @val decimal(12,2) = 0.06;
	declare @epsilon decimal(12,2) = 0.01;
	declare @lowerbound decimal(12,2), @upperbound decimal(12,2);
	set @lowerbound = @val - @epsilon;
	set @upperbound = @val + @epsilon;

	if(@discount >= @lowerbound AND @discount <= @upperbound)
		return 1;

	return 0;
end
\end{Verbatim}
}
\divider

{\scriptsize
\begin{Verbatim}[obeytabs, tabsize=2]
 create function dbo.q7conditions(@n1name varchar(25), 
																	@n2name varchar(25), 
																	@shipdate date) 
returns int as 
 begin
	if(@shipdate NOT BETWEEN '1995-01-01' AND '1996-12-31')
		return 0;

	if(@n1name = 'FRANCE' AND @n2name = 'GERMANY')
		return 1;
	else if(@n1name = 'GERMANY' AND @n2name = 'FRANCE')
		return 1;

	return 0;
 end
\end{Verbatim}
}
\divider

{\scriptsize
\begin{Verbatim}[obeytabs, tabsize=2]
create function dbo.q10conditions(@odate date, @retflag char(1)) 
returns int as 
begin
	declare @stdatechar varchar(10) = '1993-10-01';
	declare @stdate date = cast(@stdatechar as date);
	declare @newdate date = dateadd(mm, 3, @stdate);
	
	if(@retflag <> 'R')
		return 0;
	if(@odate >= @stdatechar AND @odate < @newdate) 
		return 1;

	return 0;
end
\end{Verbatim}
}
\divider

{\scriptsize
\begin{Verbatim}[obeytabs, tabsize=2]
create function dbo.total_value() returns decimal(12,2) as 
begin
	return (SELECT SUM(PS_SUPPLYCOST*PS_AVAILQTY) * 0.0001000000
			FROM PARTSUPP, SUPPLIER, NATION
			WHERE PS_SUPPKEY = S_SUPPKEY 
			AND S_NATIONKEY = N_NATIONKEY AND N_NAME = 'GERMANY');	
end 
\end{Verbatim}
}
\divider

{\scriptsize
\begin{Verbatim}[obeytabs, tabsize=2]
create function dbo.line_count(@oprio char(15), @mode varchar(4)) 
returns int as 
begin
	declare @val int = 0;
	if(@mode = 'high')
	begin
		if(@oprio = '1-URGENT' OR @oprio = '2-HIGH')
			set @val = 1;
	end
	else if(@mode = 'low')
	begin
		if(@oprio = '1-URGENT' AND @oprio = '2-HIGH')
			set @val = 1;
	end
	return @val;
end
\end{Verbatim}
}
\divider

{\scriptsize
\begin{Verbatim}[obeytabs, tabsize=2]
create function dbo.q12conditions(@shipmode char(10), 
																@commitdate date, 
																@receiptdate date, 
																@shipdate date) 
returns int as
begin
	if(@shipmode = 'MAIL' OR @shipmode ='SHIP')
	begin
		declare @stdatechar varchar(10) = '1995-09-01';
		declare @stdate date = cast(@stdatechar as date);
		declare @newdate date = dateadd(mm, 1, @stdate);

		if(@receiptdate < '1994-01-01')
			return 0;
		if(@commitdate < @receiptdate AND @shipdate < @commitdate 
				AND @receiptdate < @newdate)
			return 1;
	end	
	return 0;
end
\end{Verbatim}
}
\divider

{\scriptsize
\begin{Verbatim}[obeytabs, tabsize=2]
create function dbo.promo_disc(@ptype varchar(25), 
												@extprice decimal(12,2), 
												@disc decimal(12,2)) 
returns decimal(12,2) as
begin
	declare @val decimal(12,2);

	if(@ptype LIKE 'PROMO%%')
		 set @val = dbo.discount_price(@extprice, @disc);
	else
		set @val = 0.0;
	return @val;
end
\end{Verbatim}
}
\divider

{\scriptsize
\begin{Verbatim}[obeytabs, tabsize=2]
create function dbo.q19conditions(@pcontainer char(10), 
									@lqty int, 
									@psize int, 
									@shipmode char(10), 
									@shipinst char(25), 
									@pbrand char(10)) 
returns int as
begin
	declare @val int = 0;
	if(@shipmode IN('AIR', 'AIR REG') 
			AND @shipinst = 'DELIVER IN PERSON')
	begin
		if(@pbrand = 'Brand#12' 
			AND @pcontainer 
					IN ('SM CASE', 'SM BOX', 'SM PACK', 'SM PKG')
			AND @lqty >= 1 AND @lqty <= 1 + 10 
			AND @psize BETWEEN 1 AND 5)
				set @val = 1;

		if(@pbrand = 'Brand#23' 
			AND @pcontainer 
					IN ('MED BAG', 'MED BOX', 'MED PKG', 'MED PACK')
			AND @lqty >= 10 AND @lqty <= 10 + 10 
			AND @psize BETWEEN 1 AND 10)
				set @val = 1;

		if(@pbrand = 'Brand#34' 
			AND @pcontainer 
					IN ('LG CASE', 'LG BOX', 'LG PACK', 'LG PKG')
			AND @lqty >= 20 AND @lqty <= 20 + 10 
			AND @psize BETWEEN 1 AND 15)
				set @val = 1;
	end	
	return @val
end
\end{Verbatim}
}
\divider

{\scriptsize
\begin{Verbatim}[obeytabs, tabsize=2]
create function dbo.avg_actbal() returns decimal(12,2) as
begin
	return (SELECT AVG(C_ACCTBAL) FROM CUSTOMER 
		WHERE C_ACCTBAL > 0.00 
		AND SUBSTRING(C_PHONE,1,2) 
		IN ('13', '31', '23', '29', '30', '18', '17'));
end
\end{Verbatim}
}
\divider

\subsection{TPC-H Queries Rewritten using UDFs}

{\scriptsize
\begin{Verbatim}[obeytabs, tabsize=2]
-- Query 1
SELECT L_RETURNFLAG, L_LINESTATUS, SUM(L_QUANTITY) AS SUM_QTY,
 SUM(L_EXTENDEDPRICE) AS SUM_BASE_PRICE, 
 SUM(dbo.discount_price(L_EXTENDEDPRICE, L_DISCOUNT)) 
											AS SUM_DISC_PRICE,
 SUM(dbo.discount_taxprice(L_EXTENDEDPRICE, L_DISCOUNT, L_TAX)) 
												AS SUM_CHARGE, 
 AVG(L_QUANTITY) AS AVG_QTY,
 AVG(L_EXTENDEDPRICE) AS AVG_PRICE, AVG(L_DISCOUNT) AS AVG_DISC, 
 COUNT(*) AS COUNT_ORDER
FROM LINEITEM
WHERE dbo.isShippedBefore(L_SHIPDATE, -90, '1998-12-01') = 1
GROUP BY L_RETURNFLAG, L_LINESTATUS
ORDER BY L_RETURNFLAG,L_LINESTATUS
\end{Verbatim}
}
\divider

{\scriptsize
\begin{Verbatim}[obeytabs, tabsize=2]
-- Query 3
SELECT TOP 10 L_ORDERKEY, 
  SUM(dbo.discount_price(L_EXTENDEDPRICE, L_DISCOUNT)) AS REVENUE, 
  O_ORDERDATE, O_SHIPPRIORITY
FROM CUSTOMER, ORDERS, LINEITEM
WHERE C_CUSTKEY = O_CUSTKEY AND L_ORDERKEY = O_ORDERKEY 
AND dbo.q3conditions(C_MKTSEGMENT, O_ORDERDATE, L_SHIPDATE) = 1
GROUP BY L_ORDERKEY, O_ORDERDATE, O_SHIPPRIORITY
ORDER BY REVENUE DESC, O_ORDERDATE
\end{Verbatim}
}
\divider

{\scriptsize
\begin{Verbatim}[obeytabs, tabsize=2]
-- Query 5
SELECT N_NAME, 
SUM(dbo.discount_price(L_EXTENDEDPRICE, L_DISCOUNT)) AS REVENUE
FROM CUSTOMER, ORDERS, LINEITEM, SUPPLIER, NATION, REGION
WHERE C_CUSTKEY = O_CUSTKEY AND L_ORDERKEY = O_ORDERKEY 
AND L_SUPPKEY = S_SUPPKEY AND C_NATIONKEY = S_NATIONKEY 
AND S_NATIONKEY = N_NATIONKEY AND N_REGIONKEY = R_REGIONKEY
AND dbo.q5Conditions(R_NAME, O_ORDERDATE) = 1
GROUP BY N_NAME
ORDER BY REVENUE DESC
\end{Verbatim}
}
\divider

{\scriptsize
\begin{Verbatim}[obeytabs, tabsize=2]
-- Query 6
SELECT SUM(L_EXTENDEDPRICE*L_DISCOUNT) AS REVENUE
FROM LINEITEM
WHERE dbo.q6conditions(L_SHIPDATE, L_DISCOUNT, L_QUANTITY) = 1;
\end{Verbatim}
}
\divider

{\scriptsize
\begin{Verbatim}[obeytabs, tabsize=2]
-- Query 7
SELECT SUPP_NATION, CUST_NATION, L_YEAR, SUM(VOLUME) AS REVENUE
FROM ( SELECT N1.N_NAME AS SUPP_NATION, N2.N_NAME AS CUST_NATION, 
 datepart(yy, L_SHIPDATE) AS L_YEAR,
 L_EXTENDEDPRICE*(1-L_DISCOUNT) AS VOLUME
 FROM SUPPLIER, LINEITEM, ORDERS, CUSTOMER, NATION N1, NATION N2
 WHERE S_SUPPKEY = L_SUPPKEY AND O_ORDERKEY = L_ORDERKEY 
 AND C_CUSTKEY = O_CUSTKEY
 AND S_NATIONKEY = N1.N_NATIONKEY 
 AND C_NATIONKEY = N2.N_NATIONKEY 
 AND dbo.q7conditions(N1.N_NAME, N2.N_NAME, L_SHIPDATE) = 1 ) 
										AS SHIPPING
GROUP BY SUPP_NATION, CUST_NATION, L_YEAR
ORDER BY SUPP_NATION, CUST_NATION, L_YEAR
\end{Verbatim}
}
\divider

{\scriptsize
\begin{Verbatim}[obeytabs, tabsize=2]
-- Query 9
SELECT NATION, O_YEAR, SUM(AMOUNT) AS SUM_PROFIT
FROM (SELECT N_NAME AS NATION, 
 datepart(yy, O_ORDERDATE) AS O_YEAR,
 dbo.profit_amount(L_EXTENDEDPRICE, L_DISCOUNT, PS_SUPPLYCOST, L_QUANTITY) 
												AS AMOUNT
 FROM PART, SUPPLIER, LINEITEM, PARTSUPP, ORDERS, NATION
 WHERE S_SUPPKEY = L_SUPPKEY AND PS_SUPPKEY= L_SUPPKEY 
 AND PS_PARTKEY = L_PARTKEY AND P_PARTKEY= L_PARTKEY 
 AND O_ORDERKEY = L_ORDERKEY AND S_NATIONKEY = N_NATIONKEY AND
 P_NAME LIKE '%%green%%') AS PROFIT
GROUP BY NATION, O_YEAR
ORDER BY NATION, O_YEAR DESC
\end{Verbatim}
}
\divider

{\scriptsize
\begin{Verbatim}[obeytabs, tabsize=2]
-- Query 10
SELECT TOP 20 C_CUSTKEY, C_NAME, 
 SUM(dbo.discount_price(L_EXTENDEDPRICE, L_DISCOUNT)) AS REVENUE, 
 C_ACCTBAL, N_NAME, C_ADDRESS, C_PHONE, C_COMMENT
FROM CUSTOMER, ORDERS, LINEITEM, NATION
WHERE C_CUSTKEY = O_CUSTKEY AND L_ORDERKEY = O_ORDERKEY AND 
dbo.q10conditions(O_ORDERDATE, L_RETURNFLAG) = 1
AND C_NATIONKEY = N_NATIONKEY
GROUP BY C_CUSTKEY, C_NAME, C_ACCTBAL, C_PHONE, 
					N_NAME, C_ADDRESS, C_COMMENT
ORDER BY REVENUE DESC
\end{Verbatim}
}
\divider

{\scriptsize
\begin{Verbatim}[obeytabs, tabsize=2]
-- Query 11
SELECT PS_PARTKEY, SUM(PS_SUPPLYCOST*PS_AVAILQTY) AS VALUE
FROM PARTSUPP, SUPPLIER, NATION
WHERE PS_SUPPKEY = S_SUPPKEY AND S_NATIONKEY = N_NATIONKEY 
AND N_NAME = 'GERMANY'
GROUP BY PS_PARTKEY
HAVING SUM(PS_SUPPLYCOST*PS_AVAILQTY) > dbo.total_value()
ORDER BY VALUE DESC
\end{Verbatim}
}
\divider

{\scriptsize
\begin{Verbatim}[obeytabs, tabsize=2]
-- Query 12
SELECT L_SHIPMODE,
SUM(dbo.line_count(O_ORDERPRIORITY, 'high')) AS HIGH_LINE_COUNT,
SUM(dbo.line_count(O_ORDERPRIORITY, 'low')) AS LOW_LINE_COUNT
FROM ORDERS, LINEITEM
WHERE O_ORDERKEY = L_ORDERKEY AND 
dbo.q12conditions(L_SHIPMODE, L_COMMITDATE, 
					L_RECEIPTDATE, L_SHIPDATE) = 1
GROUP BY L_SHIPMODE
ORDER BY L_SHIPMODE
\end{Verbatim}
}
\divider

{\scriptsize
\begin{Verbatim}[obeytabs, tabsize=2]
-- Query 14
SELECT 100.00 * 
	SUM(dbo.promo_disc(P_TYPE, L_EXTENDEDPRICE, L_DISCOUNT)) 
	/ SUM(dbo.discount_price(L_EXTENDEDPRICE,L_DISCOUNT)) 
				AS PROMO_REVENUE
FROM LINEITEM, PART
WHERE L_PARTKEY = P_PARTKEY AND L_SHIPDATE >= '1995-09-01' 
AND L_SHIPDATE < dateadd(mm, 1, '1995-09-01')
\end{Verbatim}
}
\divider

{\scriptsize
\begin{Verbatim}[obeytabs, tabsize=2]
-- Query 19
SELECT SUM(dbo.discount_price(L_EXTENDEDPRICE, L_DISCOUNT)) 
			AS REVENUE
FROM LINEITEM join PART on L_PARTKEY = P_PARTKEY
WHERE dbo.q19conditions(P_CONTAINER, L_QUANTITY, P_SIZE, 
			L_SHIPMODE, L_SHIPINSTRUCT, P_BRAND ) = 1;
\end{Verbatim}
}
\divider

{\scriptsize
\begin{Verbatim}[obeytabs, tabsize=2]
-- Query 22
SELECT CNTRYCODE, 
		COUNT(*) AS NUMCUST, SUM(C_ACCTBAL) AS TOTACCTBAL
FROM (SELECT SUBSTRING(C_PHONE,1,2) AS CNTRYCODE, C_ACCTBAL
 FROM CUSTOMER WHERE SUBSTRING(C_PHONE,1,2) 
 			IN ('13', '31', '23', '29', '30', '18', '17') 
		AND C_ACCTBAL > dbo.avg_actbal() 
		AND NOT EXISTS ( SELECT * FROM ORDERS 
				WHERE O_CUSTKEY = C_CUSTKEY)) AS CUSTSALE
GROUP BY CNTRYCODE
ORDER BY CNTRYCODE
\end{Verbatim}
}
\divider

\section{Related Work} \label{sec:relwork}
Optimization of SQL queries containing sub-queries is well-studied. There have been several techniques proposed over 
the years~\cite{KIM82,GAN87,SPL96,Gal01,DAYAL87,GAL07, NeumannUnnest}, and many RDBMSs 
can optimize nested sub-queries. Complementarily, there has been a lot 
of work spanning multiple decades, on optimization of imperative programs in the 
compilers community~\cite{Aho06, MUCHNICK, KENNEDYBOOK}. UDFs are similar 
to nested sub-queries, but contain imperative constructs. Hence, they lie in 
the intersection of these two streams of work; however, they have received little attention 
from either community.

Some databases perform sub-program inlining, which applies only to nested function 
calls~\cite{ORASPI}. This technique works by replacing the call to a function 
with the function body. 
%This reduces invocation overheads but does not overcome
%any of the limitations listed in \refsec{sec:udfperf}. 
Another technique is to cache function results~\cite{ResCache}, which is useful only 
when there are repeated UDF invocations with identical parameter values. 
%While these are partial solutions, Froid is a complete solution 
%that addresses all drawbacks of UDF evaluation.
Unlike Froid, none of these techniques offer a complete solution that addresses 
all drawbacks of UDF evaluation listed in \refsec{sec:udfperf}. 

%There have been recent efforts that use programming languages
%techniques to optimize database-backed applications. 
%In particular, the work of
%Cheung et al.~\cite{Cheung13, CHEUNGCIDR13} and Emani et al.~\cite{Emani2016} focus on generating
%equivalent SQL from imperative code. While Cheung et al.~\cite{Cheung13} use program 
%synthesis to infer SQL queries, Emani et al.~\cite{Emani2016} present a static 
%analysis based approach with similar goals. 

There have been recent efforts that use programming languages
techniques to optimize database-backed applications. 
Cheung et al.~\cite{Cheung13} consider applications 
written using object-relational mapping libraries and transforms fragments
of code into SQL using Query-By-Synthesis (QBS).
The goals of QBS and Froid are similar, but the approaches are entirely
different. QBS is based on program synthesis, whereas Froid uses a program
transformation based approach.
Although QBS is a powerful technique, it is limited in its scalability to 
large functions. We have manually 
analyzed all code fragments used in~\cite{Cheung13} (given in Appendix A of~\cite{Cheung13}), and found that 
none of those are larger than 100 lines of code. Even for these small code 
fragments, QBS suffers from potentially very long optimization times  
due to the space-exploration involved. They use a preset timeout of 10 mins 
in their experiments. Froid overcomes both these limitations -- 
it can handle UDFs with 1000s of statements, and can transform them in less 
than 10 seconds (see~\refsec{subsubsec:comptime}).

The StatusQuo system~\cite{CHEUNGCIDR13} includes (a) a program analysis that 
identifies blocks of imperative logic that can be translated to SQL and 
(b) a program partitioning method to move application logic into imperative stored procedures. 
The SQL translation in StatusQuo uses QBS\cite{Cheung13} to extract equivalent SQL. 
The program partitioning is orthogonal to our work. Once such 
partitioning is done, the resulting imperative procedures can be optimized using Froid.

%- cases where sql is embedded in imperative code	

%Although the technique of \cite{Cheung13} is quite powerful in identifying code 
%fragments that can be replaced by SQL, their approach is resource intensive, and
%is not suitable for integration into a database engine. 
Emani et al.~\cite{Emani2016, Emani2017} present a static analysis based approach with similar goals.
and can be adapted to extract equivalent SQL for UDFs. However, 
from a prototype implementation, we found that the generated SQL turns out to be 
larger and more complex compared to Froid. As discussed in \refsec{subsec:limits}, 
we prefer to minimize the size of input to the optimizer.

Simhadri et al.~\cite{uudf} describe a technique to decorrelate queries in UDFs 
using extensions to the \textit{Apply} operator. Froid's approach
borrows its intuition from this work, but there are some key differences.
First, 
Froid does not require any new operators or operator extensions unlike 
%where\-as 
the approach of~\cite{uudf}.
%is based on their extensions to the \textit{Apply} operator. 
Second, their transformation rules are designed to be a part of a cost 
based optimizer. Froid, in contrast is designed as a precursor to query 
optimization, and requires no modifications to the query optimizer.
%, and does not perform cost based substitution. 
We argue that inlining need not be cost-based, but 
should instead be treated like view substitution (\refsec{subsec:cbi}).
Third, they do not address vital issues such as handling multiple return statements 
and avoiding redundant computation of predicate expressions, which are found 
to be quite common in real workloads. 
We describe the design of a complete 
optimization framework that is extensible to other languages, and also show 
how compiler optimizations can be expressed as relational transformations.
\section{Conclusion} \label{sec:concl}
While declarative SQL and procedural extensions are both supported by RDBMSs, 
their primary focus has been the efficient evaluation of 
declarative SQL. 
%The poor performance of UDFs is a
%major concern, and using UDFs is often discouraged for this reason.
Although imperative UDFs and procedures offer many 
advantages and are preferred by many users, their poor performance is
a major concern. Often, using UDFs is discouraged for this reason.

In this paper, we address this important problem using novel techniques that 
automatically transform imperative programs into relational expressions. This 
enables us to leverage sophisticated query optimization techniques thereby resulting in
efficient, set-oriented, parallel plans for queries
invoking UDFs. Froid, our extensible, language-agnostic optimization framework built 
into Microsoft SQL Server, not only overcomes current drawbacks in UDF 
evaluation, but also offers the benefits of many compiler optimization 
techniques with no additional effort. 
The benefits of our framework are dem\-onstrated by our %extensive
evaluation on customer workloads, showing significant gains. 
We believe that our work will enable and encourage the wider use of UDFs to build 
modular, reusable and maintainable applications without compromising 
performance.

%As more complex applications are built, % using these systems, 
%the need to efficiently evaluate procedural code in an RDBMS is becoming 
%increasingly important. 

%In this paper, we presented Froid, a framework to optimize T-SQL UDFs and
%procedures in Microsoft SQL Server. 

%We showed how Froid incorporates ideas from programming 
%language compilers and relational query optimizers and is able to overcome 
%limitations in procedural code evaluation in databases. 

%To the best 
%of our knowledge, Froid is the first industrial strength framework that 
%integrates compiler optimization techniques into a RDBMS for optimization of 
%procedural code. The benefits of such a framework are clear 
%from our extensive evaluation on real world customer workloads.

% that shows improvements of up to a factor of 
%1000 due to our framework.
%We have conducted an extensive evaluation of Froid on real world customer workloads
%that show the significant gains of such a framework. 

%\end{document}  % This is where a 'short' article might terminate

% ensure same length columns on last page (might need two sub-sequent latex runs)
\balance

%ACKNOWLEDGMENTS are optional
\section{Acknowledgments}
The authors would like to thank David DeWitt, Jignesh Patel and Mike Zwilling
for their support and feedback.

\bibliographystyle{abbrv}
\small{\bibliography{froid-tr}}  % vldb_sample.bib is the name of the Bibliography in this case

%\subsection{References}
%Generated by bibtex from your ~.bib file.  Run latex,
%then bibtex, then latex twice (to resolve references).

%APPENDIX is optional.
% ****************** APPENDIX **************************************
% Example of an appendix; typically would start on a new page
%pagebreak

%\input{appendix.tex}

\end{document}